\documentclass[useAMS,usenatbib,usegraphicx]{mn2e}
\usepackage{graphicx}
\usepackage{graphics}
\usepackage{color}
\usepackage{float}
\usepackage{longtable}
\usepackage{appendix}
\usepackage{mdwlist} 
\usepackage{lscape}
\usepackage{paralist}

\DeclareGraphicsExtensions{.pdf}

\begin{document}

\def\ymgas{$Y_{500}-M_{\rm{gas}}\,$}
\def\ytx{$Y_{500}-T_X\,$}
\def\yyx{$Y_{500}-Y_X\,$}
\def\y500{$Y_{500}$}
\def\arcdeg{\mbox{$^\circ$}}%
\def\arcmin{\mbox{$^\prime$}}
\def\arcsec{\mbox{$^{\prime\prime}$}}
\newcommand\nodata{ ~$\cdots$~ }%

\def\jref@jnl#1{{#1}}%
\newcommand\apj{\jref@jnl{ApJ}}%
\newcommand\apjl{\jref@jnl{ApJ}}%
\newcommand\apjs{\jref@jnl{ApJS}}%
\newcommand\mnras{\jref@jnl{MNRAS}}%
\newcommand\aap{\jref@jnl{A\&A}}%
\newcommand\aj{\jref@jnl{AJ}}%
\newcommand\icarus{\jref@jnl{Icarus}}%
\newcommand\araa{\jref@jnl{ARA\&A}}%
\newcommand\pasp{\jref@jnl{PASP}}%
\newcommand\jcap{\jref@jnl{J. Cosmology Astropart. Phys.}}%

\def\McGill{$^1$ }
\def\Cardiff{$^2$ }
\def\Bonn{$^3$ }
\def\Berkeley{$^4$ }
\def\LBNLMS{$^5$ }
\def\OSO{$^6$ }
\def\inao{$^7$ }
\def\CUA{$^{8}$}
\def\CUP{$^9$ }
\def\princeton{$^{10}$ }
\def\ESO{$^{11}$}
\def\ALMA{$^{12}$ }
\def\LBNL{$^{13}$ }
\def\MPI{$^{14}$ }


\title{Galaxy Cluster Scaling Relations Measured with APEX-SZ}

\author[A.~N.~Bender et al.]{A.~N.~Bender,$^1$\thanks{Corresponding author email: abender@anl.gov}
J.~Kennedy,\McGill P.~A.~R.~Ade,\Cardiff  K.~Basu,\Bonn  F.~Bertoldi,\Bonn  
\newauthor
S.~Burkutean,\Bonn  J.~Clarke,$^{4,}$\LBNLMS D.~Dahlin,\OSO M.~Dobbs,\McGill 
D.~Ferrusca,\inao 
\newauthor
D.~Flanigan,\Berkeley 
N.~W.~Halverson,\CUA$^{,}$\CUP 
W.~L.~Holzapfel,\Berkeley 
C.~Horellou,\OSO 
\newauthor
B.~R.~Johnson,\Berkeley 
Z.~D.~Kermish,\princeton 
M.~Klein,\Bonn 
R.~Kneissl,\ESO$^{,}$\ALMA 
T.~Lanting,\McGill 
\newauthor
A.~T.~Lee,$^{4,}$\LBNL 
J.~Mehl,\Berkeley 
K.~M.~Menten,\MPI 
D.~Muders,\MPI 
A.~Nagarajan,\Bonn
\newauthor 
F.~Pacaud,\Bonn 
C.~L.~Reichardt,\Berkeley 
P.~L.~Richards,\Berkeley 
R.~Schaaf,\Bonn 
D.~Schwan,\Berkeley
\newauthor 
M.~W.~Sommer,\Bonn 
H.~Spieler,\LBNL 
C.~Tucker,\Cardiff 
B.~Westbrook\Berkeley\\
\McGill Department of Physics, McGill University, Montr\'{e}al,Canada, H3A 2T8\\
\Cardiff School of Physics and Astronomy, Cardiff University,CF24 3YB Wales, UK\\
\Bonn Argelander Institute for Astronomy, Bonn University, Bonn, Germany\\
\Berkeley Department of Physics, University of California, Berkeley, CA, 94720\\
\LBNLMS Materials Sciences Division, Lawrence Berkeley National Laboratory, Berkeley, CA, 94720\\
\OSO Dept. of Earth \& Space Sciences, Chalmers University of Technology, Onsala Space Observatory, SE-439 92 Onsala, Sweden\\
\inao Instituto Nacional de Astrof\'{i}sica, \'{O}ptica y Electr\'{o}nica, Luis Enrique Erro 1, Tonantzintla, Puebla, C.P. 72840 M\'{e}xico\\
\CUA Center for Astrophysics and Space Astronomy, Department of Astophysical and Planetary Sciences, University of Colorado, \\      ~~Boulder, CO, 80309\\
\CUP Department of Physics, University of Colorado,Boulder, CO, 80309\\
\princeton Department of Physics, Princeton University, Princeton, NJ, 08544\\
\ESO European Southern Observatory, Alonso de C\'{o}rdova 3107, Vitacura, Santiago, Chile\\
\ALMA Atacama Large Millimeter/submillimeter Array, Joint ALMA Observatory, Alonso de C\'{o}rdova 3107, Vitacura, Santiago, Chile \\
\LBNL Physics Division, Lawrence Berkeley National Laboratory, Berkeley, CA, 94720\\
\MPI Max Planck Institute for Radio Astronomy, 53121 Bonn, Germany}

\maketitle

\begin{abstract}
We present thermal Sunyaev-Zel'dovich effect (SZE) measurements for 42 galaxy clusters observed at 150 GHz with the APEX-SZ experiment.   
For each cluster, we model the pressure profile and calculate the integrated Comptonization $Y$ to estimate the total thermal energy of the intracluster medium (ICM).    
We compare the measured $Y$ values to X-ray observables of the ICM from the literature (cluster gas mass $M_{\rm{gas}}$, temperature $T_X$, and $Y_X =M_{\rm{gas}}T_X$) that relate to total cluster mass.  We measure power law scaling relations, including an intrinsic scatter, between the SZE and X-ray observables for three subsamples within the set of 42 clusters that have uniform X-ray analysis in the literature.  We observe that differences between these X-ray analyses introduce significant variability into the measured scaling relations, particularly affecting the normalization.  For all three subsamples, we  find results consistent with a self-similar model of cluster evolution dominated by gravitational effects.  Comparing to predictions from numerical simulations, these scaling relations prefer models that include cooling and feedback in the ICM. Lastly, we measure an intrinsic scatter of $\sim28$ per cent in the $Y-Y_X\,$ scaling relation for all three subsamples.

\end{abstract}



\section{Introduction}
\label{SEC:introduction}

\setcounter{footnote}{0}
As the largest gravitationally collapsed objects in the Universe, clusters of galaxies provide a unique opportunity to study the evolution of large-scale structure.  The distribution and abundance of clusters is sensitive to both the geometry of the universe and the growth of density perturbations \citep[e.g.,][]{haiman2001,weller2002}.   Currently, cluster-based constraints on cosmology are limited by systematic uncertainties in relating observables to cluster masses.
Most of the cluster mass is in the form of dark matter and therefore is not directly observable.  Instead, cluster masses are inferred through scaling relations with observable signals such as  X-ray luminosity, galaxy velocity distribution,  weak-lensing shear and Sunyaev-Zel'dovich effect (SZE) brightness.

Under the model of self-similarity, where cluster evolution is dominated by gravitational processes, cluster mass scales with observable signals through simple power law relations \citep{kaiser1986}.   In this model, the intracluster medium (ICM) is in hydrostatic equilibrium, for which the scaling to cluster mass can be predicted for a given observable.  However, self-similarity does not take into account the role of non-thermal mechanisms such as turbulent gas motions in estimating cluster mass.  This simple model also neglects the effects of physical processes internal to the cluster such as feedback from active galactic nuclei and star formation.  Numerical simulations predict that while the power law exponent of SZE scaling relations will be consistent with self-similarity, the normalization does depend on the internal cluster astrophysics \citep{motl2005, nagai2006, lau2009}.  An empirical measurement of the scaling relations therefore informs models of cluster astrophysics, progressing towards the needed calibration for cosmology.
 
Currently, cluster surveys are operating in the optical \citep[e.g.,][]{gilbank2011}, X-ray \citep[e.g.,][]{vikhlinin2009,mantz2010}, and the millimeter \citep[e.g.,][]{carlstrom2011,swetz2011,planck2013xxix} wavelength regimes.  However, optical and X-ray measurements of clusters suffer from cosmological dimming, and only the brightest and most massive clusters are detected at high redshifts ($z > 1$).  In contrast, the millimeter-wavelength thermal SZE \citep{sunyaev72}, where Cosmic Microwave Background (CMB) photons inverse-Compton scatter off hot intracluster electrons, is redshift-independent.  
SZE surveys with sufficient resolution to resolve clusters, such as those performed with the South Pole Telescope \citep{reichardt2013} and the Atacama Cosmology Telescope \citep{hasselfield2013} detect clusters with a mass selection nearly independent of redshift.

Precision cosmology requires that mass-observable scaling relations be characterized with a high level of accuracy, including both the measurement uncertainty and the intrinsic scatter from cluster-to-cluster differences. Numerical simulations suggest that SZE observations are relatively insensitive to the details of cluster astrophysics \citep[e.g.,][]{hallman2006, nagai2006}, resulting in low intrinsic scatter scaling relations and a tighter constraint on cluster mass.   Previous measurements of SZE scaling relations include interferometric observations from the OVRO/BIMA \citep{bonamente2008} and SZA \citep{culverhouse2010, marrone2012} arrays as well as imaging studies with the South Pole Telescope \citep{andersson2011,plagge2010,benson2013}, BOLOCAM \citep{sayers2011}, Atacama Cosmology Telescope \citep{marriage2010}, and the Planck mission \citep{planck2011XI}.  In general, those studies find that observational SZE scaling relations agree with expectations based on self-similarity and that simulations including additional non-adiabatic physics are preferred.

While cluster samples selected from large surveys (both SZE and X-ray) have well-known selection functions, studies such as the one presented in this paper that target known clusters often select them in an \textit{ad hoc} manner.  
The influence of this sample selection on SZE scaling relations is unknown.

In this paper, we study how the SZE signal scales with X-ray observables related to cluster mass using observations from the APEX-SZ imaging bolometer array.  We measure SZE scaling relations using X-ray observables ($Y_X$, $M_{\rm{gas}}$, and $T_X$) as proxies for the total cluster mass.   APEX-SZ observed a small set of 11 clusters selected from the REFLEX X-ray survey (referred to as the REFLEX-DXL sample \citep{zhang2006}) as well as an additional 31 clusters selected in an \textit{ad hoc} manner.   Within this full sample are clusters drawn from the \citet[][hereafter Z08]{zhang2008} and \citet[][hereafter M10]{mantz2010} samples.  We measure power law scaling relations for each of these three subsamples (REFLEX-DXL, Z08, M10) and compare the results to the expectations of the self-similar model as well as numerical simulations that incorporate different physical processes in the ICM.  We compare the results between the three  cluster samples as well as to the full APEX-SZ sample to investigate the effects of sample selection and uniform analysis on the measured power law and intrinsic scatter of the  SZE scaling relations.  
In this paper, we assume the \textit{WMAP7}+BAO+$H_0$ $\Lambda$CDM best-fit cosmology in which $H_0=70.4$ km\,$\rm{s}^{-1}$\,$\rm{Mpc}^{-1}$, $\Omega_M=0.272$, and $\Omega_{\Lambda}=0.728$ \citep{komatsu2011}.  

The structure of the paper is as follows.  Section~\ref{sec:theory} reviews the SZE and the associated scaling relations.  Section \ref{SEC:observations} introduces the clusters in this study and observations thereof.  Section \ref{SEC:szdatareduction} briefly describes the data reduction process for the APEX-SZ instrument.  We present the  SZE-X-ray scaling relations and discuss these results in Section~\ref{SEC:results}.  A summary and an outlook on future work are provided in Section \ref{SEC:conclusions}.
\section{The Thermal Sunyaev-Zel'dovich Effect}
\label{sec:theory}

The SZE occurs when photons from the CMB  inverse Compton scatter off hot electrons in the intracluster medium to higher energies \citep{sunyaev72}.  This process distorts the CMB blackbody spectrum with an amplitude proportional to the gas pressure integrated along the line-of-sight.  In units of temperature the change is
\begin{equation}
\label{EQN:szeffect}
\frac{\Delta T_{\rm{SZE}}}{T_{\rm{CMB}}} = \int f(x,T_e)\cdot n_e \sigma_T\frac{k_BT_e}{m_ec^2} dl \equiv f(x,T_e) \cdot y,
\end{equation}
where $n_e$ and $T_e$ are the electron density and temperature of the ICM, $T_{\rm{CMB}}$ is the temperature of the CMB, $\sigma_T$ is the Thomson cross-section, $k_B$ is the Boltzmann constant, and $m_ec^2$ is the electron rest mass energy.  The integrated pressure is often parametrized in terms of the Comptonization, $y$.  The frequency dependence $f(x,T_e)$  is given by
\begin{equation}
\label{EQN:szefreq}
f(x,T_e)=\left(x\frac{e^x+1}{e^x-1}-4\right)(1+\delta_{\rm{SZE}}(x,T_e)),
\end{equation}
where $x=h\nu/k_BT_{\rm{CMB}}$ and $\delta_{\rm{SZE}}(x,T_e)$ is a correction for relativistic electrons \citep[see, e.g., ][]{nozawa2000}. 
For frequencies below 217 GHz, the SZE manifests itself as a decrement in the CMB spectrum, while at frequencies above 217 GHz the SZE results in an increment.  Equation \ref{EQN:szeffect} demonstrates two important characteristics of the SZE.  First, the SZE is redshift independent since its surface brightness is expressed as a fractional change in the CMB temperature.  Second, the dependence on integrated pressure directly measures the thermal energy of the ICM.

The Comptonization integrated over the solid angle of the cluster on the sky, 
\begin{equation}
\label{EQN:szint}
Y= \int y d\Omega,
\end{equation}
 measures the total thermal energy of a cluster, and is therefore expected to be a robust proxy for total mass.  In this work, we integrate a parametric model of $Y$ to a radius of $r_{500}$, the radius at which the mean matter density of the cluster is 500 times the critical density of the Universe, $\rho_{c}(z)=3H_0^2E(z)^2/8\pi G$.  
Assuming that clusters are in hydrostatic equilibrium, the integrated Comptonization scales as a function of the ICM properties according to
\begin{eqnarray}
\label{EQN:scale}
Yd_A^2 &\propto& f_{\rm{gas}}T_e^{5/2}E(z)^{-1}\nonumber\\
Yd_{A}^2 &\propto& f_{\rm{gas}}^{-2/3}M_{\rm{gas}}^{5/3}E(z)^{2/3},
\end{eqnarray}
where $E(z)$ is the expansion rate of the universe normalized to its present value, $E(z) = H(z)/H_0= \sqrt{\Omega_{M}(1+z)^3 + \Omega_{\Lambda}}$, $f_{\rm{gas}}$ is the gas mass fraction, and $d_A$ is the angular diameter distance of the cluster \citep{bonamente2008}.   A third proxy for cluster mass can be constructed from the combination of temperature and gas mass, $Y_X = M_{\rm{gas}}T_X$  \citep{kravtsov2006}.   $Y$ scales with $Y_X$ as
\begin{equation}
\label{EQN:yyxscale}
Yd_A^2  \propto Y_X,
\end{equation}
and is expected to have lower intrinsic scatter than $Y-T$ or $Y-M_{\rm{gas}}$ due to the anti-correlation of systematic deviations between $T$ and $M_{\rm{gas}}$.

\section{Observations}
\label{SEC:observations}

APEX-SZ  \citep{schwan2011} was a transition-edge-sensor (TES) bolometer array located on the 12-meter Atacama Pathfinder Experiment (APEX) telescope in northern Chile \citep{2006SPIE.6267E..37G}.   The focal plane comprised a total of 280 bolometers distributed on six wafers which were cooled with closed-cycle refrigerators to an average operating temperature of $\sim$300 mK and read out using frequency domain multiplexing \citep{dobbs2012}.  The APEX-SZ experiment imaged SZE decrements at 150 GHz with arcmin resolution over a 23 arcmin field of view (FOV).   Additional details of the APEX-SZ instrument are presented in \citet{dobbs2006} and \citet{schwan2011}.  

APEX-SZ completed two commissioning and seven observing runs from 2005 to 2010.  During this time, APEX-SZ observed the 42 clusters presented here for a total of approximately 760 hours.  APEX-SZ executed a circular drift scan pattern that concentrated the integration time in a region around the known cluster coordinates while limiting overhead time due to telescope turn-arounds.   The telescope repeated the circular scan while centered on a constant azimuth and elevation, allowing the target to drift through the FOV.  After completing between 11 and 20 circles (4--7 seconds per circle), the telescope slewed to track the target and then repeated the circular pattern.  The radius of each circle ($6$--$12$ arcmin) was chosen from the expected radial extent of each cluster to ensure that the detectors observed the sky background during each scan.  These scans resulted in a sky coverage area of approximately $0.75\arcdeg \times 1.0 \arcdeg$.  

Clusters were targeted based on the availability of published X-ray measurements.  In general, massive clusters ($T_X > 6$ keV) were selected and the sample included both dynamically relaxed and disturbed clusters.  Additionally, APEX-SZ targeted clusters from the REFLEX-DXL X-ray sample \citep{zhang2006}, which is discussed in further detail in \ref{SEC:sampleintro}.  Coordinates and map depth for each cluster are given in Table \ref{TBL:clustercoord}.

\begin{table*}
\begin{minipage}{126mm}
\begin{center}
\caption{Summary of cluster properties.}
\label{TBL:clustercoord}
\begin{tabular}{lcccc}
\hline\vspace{-2.5ex}\\
Cluster Name & Right Ascension &  Declination & redshift  & Map Depth$^{\rm{a}}$\\
 &(\textit{h m s})& ({\arcdeg} {\arcmin} {\arcsec})  & &($\mu K$ CMB)\\
\hline\vspace{-2.5ex}\\
Abell 2744 &00 14 18.6 & -30 23 15.4 &  0.307 &   12 \\ 
Abell 2813 &00 43 24.5 & -20 37 31.2 &  0.292 &   21 \\ 
Abell 209 &01 31 52.6 & -13 36 35.5 &  0.209 &   16 \\ 
XLSS J022145.2-034617 &02 21 45.2 & -03 46 17.4 &  0.430 &    7 \\ 
RXCJ0232.2-4420 &02 32 18.8 & -44 20 51.9 &  0.284 &   17 \\ 
Abell 383 &02 48 03.3 & -03 31 43.6 &  0.187 &   14 \\ 
RXCJ0437.1+0043 &04 37 09.5 & +00 43 54.5 &  0.284 &   14 \\ 
MS0451.6-0305 &04 54 11.3 & -03 00 52.6 &  0.550 &   21 \\ 
Abell 520 &04 54 09.0 & +02 55 18.0 &  0.203 &   14 \\ 
RXCJ0516.6-5430 &05 16 35.2 & -54 30 36.8 &  0.294 &   10 \\ 
RXCJ0528.9-3927 &05 28 52.5 & -39 28 16.7 &  0.284 &   11 \\ 
RXCJ0532.9-3701 &05 32 55.9 & -37 01 34.5 &  0.275 &   16 \\ 
Abell 3404 &06 45 30.0 & -54 13 42.1 &  0.164 &   13 \\ 
1ES 0657-56 &06 58 30.2 & -55 56 33.7 &  0.296 &   19 \\ 
Abell 907 &09 58 21.9 & -11 03 48.2 &  0.160 &   11 \\ 
XMMXCSJ095940.8+023111.3 &09 59 40.8 & +02 31 11.3 &  0.720 &   14 \\ 
RXCJ1023.6+0411 &10 23 39.6 & +04 11 12.0 &  0.291 &   12 \\ 
MS1054.4-0321 &10 56 59.0 & -03 37 37.0 &  0.830 &   13 \\ 
MACSJ1115.8+0129 &11 15 51.9 & +01 29 55.0 &  0.355 &   16 \\ 
Abell 1300 &11 31 54.7 & -19 55 40.5 &  0.308 &   18 \\ 
RXCJ1206.2-0848 &12 06 12.3 & -08 48 06.0 &  0.439 &   18 \\ 
XMMUJ1230.3+1339 &12 30 16.9 & +13 39 04.3 &  0.975 &   11 \\ 
RDCSJ1252-2927 &12 52 54.4 & -29 27 17.0 &  1.240 &    8 \\ 
MACSJ1311.0-0311 &13 11 01.7 & -03 10 37.6 &  0.494 &   12 \\ 
Abell 1689 &13 11 29.3 & -01 20 26.7 &  0.184 &   19 \\ 
RXCJ1347.5-1144 &13 47 30.8 & -11 45 09.0 &  0.451 &   28 \\ 
MACSJ1359.1-1929 &13 59 10.3 & -19 29 24.0 &  0.447 &   27 \\ 
Abell 1835 &14 01 01.9 & +02 52 35.5 &  0.253 &   36 \\ 
RXCJ1504.1-0248 &15 04 07.6 & -02 48 16.0 &  0.215 &   21 \\ 
Abell 2163 &16 15 46.0 & -06 08 54.0 &  0.203 &   30 \\ 
Abell 2204 &16 32 47.1 & +05 34 32.3 &  0.152 &   10 \\ 
MACSJ1931.8-2635 &19 31 49.6 & -26 34 34.0 &  0.352 &   30 \\ 
RXCJ2011.3-5725 &20 11 27.1 & -57 25 09.8 &  0.279 &   11 \\ 
RXCJ2014.8-2430 &20 14 49.7 & -24 30 30.0 &  0.161 &   15 \\ 
MACSJ2046.0-3430 &20 46 00.5 & -34 30 17.0 &  0.423 &   15 \\ 
RXCJ2214.9-1359 &22 14 57.4 & -14 00 10.8 &  0.503 &   23 \\ 
XMMXCSJ2215.9-1738 &22 15 58.0 & -17 38 02.5 &  1.450 &   11 \\ 
XMMUJ2235.3-2557 &22 35 20.6 & -25 57 42.0 &  1.393 &    9 \\ 
RXCJ2243.3-0935 &22 43 21.4 & -09 35 43.0 &  0.447 &   20 \\ 
Abell S1077 &22 58 48.1 & -34 47 59.4 &  0.313 &   17 \\ 
Abell 2537 &23 08 22.0 & -02 11 30.0 &  0.297 &   11 \\ 
RXCJ2337.6+0016 &23 37 37.8 & +00 16 15.5 &  0.278 &   20 \\ 
\hline
\end{tabular}
\end{center}
$^{\rm{a}}$Map depth is measured within the central arcminute from the standard deviation of 500 jackknife noise map realizations.\\
\end{minipage}
\end{table*}

\begin{table*}
\begin{minipage}{126mm}
\begin{center}
\caption{Summary of cluster X-ray properties.}
\label{TBL:physparam}
\begin{tabular}{lccccc}
\hline\vspace{-3.5ex}\\
Cluster Name &$T_X$ &Ref. & $M_{\rm{gas}}$ &  Ref.&  $Y_X$  \\
  & keV &   & $10^{13} \rm{M}_{\odot}$&  & $10^{13}$keV\,$\rm{M}_{\odot}$\\
\hline\vspace{-2.5ex}\\
Abell 2744 &$10.1\pm 0.3$ & 2 & $10\pm 2$ & 2 & $101\pm20$ \\
  & $8.53\pm0.37$ &3&$20.1\pm2.1$ &3&$171.5 \pm19.4$\\ 
Abell 2813 &$7.7\pm 0.3$ & 2 & $6\pm 1$ & 2 & $46\pm8$ \\
   & $7.0\pm0.4$ & 1& $5.8\pm 0.6$&1& $40.6\pm4.8$\\
  & $7.59\pm0.77$ &3& $9.2 \pm1.4$ &3&$69.8 \pm 12.8$\\
Abell 209 &$7.1\pm 0.3$ & 1 & $7.8\pm 0.8$ & 1 & $55.4\pm6.1$\\ 
   &$8.23 \pm0.66$ &3& $14.4\pm1.8$ &3&$118.5 \pm 17.6$\\
XLSS J022145.2-034617 &$4.8_{-0.5}^{+0.6}$ &7 & \nodata &\nodata &\nodata\\ 
RXCJ0232.2-4420 &$7.0\pm 0.3$ & 2 & $9\pm 2$ & 2 & $63\pm14$ \\
    &$6.6\pm 0.3$ & 1&$8.9\pm0.9$&1&$58.7\pm6.5$\\
  &$10.06\pm 2.31$ & 3& $14.5\pm2.5$ &3& $145.9 \pm42.9$\\ 
Abell 383 &$4.7\pm 0.2$ & 1 & $3.3\pm 0.4$ & 1 & $15.5\pm2.0$ \\
RXCJ0437.1+0043 &$5.1\pm 0.3$ & 2 & $5\pm 1$ & 2 & $26\pm5$ \\
MS0451.6-0305 &$6.6_{-0.6}^{+0.7}$ &11 & $12.1_{-0.2}^{+0.2}$ & 11 & $79.7_{-7.4}^{+8.5}$\\ 
Abell 520 &$7.23\pm 0.23$ & 3 & $13.7\pm 1.5$ & 3 & $99.1\pm11.3$ \\
RXCJ0516.6-5430 &$7.5\pm 0.3$ & 2 & $8\pm 2$ & 2 & $60\pm15$ \\
    & $6.7\pm0.5$ & 1& $7.8\pm0.9$&1&$52.3 \pm7.2$\\
RXCJ0528.9-3927 &$7.2\pm 0.4$ & 2 & $9\pm 1$ & 2 & $65\pm8$ \\
   & $6.6\pm 0.5$ &1&  $8.7\pm0.7$&1& $57.4\pm6.4$\\
 &$ 7.8\pm 0.85$&3&$ 15.2 \pm1.7$&3&$118.6\pm18.5$\\
RXCJ0532.9-3701 &$9.5\pm 0.4$ & 2 & $6\pm 1$ & 2 & $57\pm10$\\
   & $7.7\pm0.6$ & 1&$5.7\pm 0.7$&1& $43.9 \pm 6.4$\\
Abell 3404 &$7.6\pm 0.3$ & 1 & $9.0\pm 1.1$ & 1 & $68.4\pm8.8$ \\
1ES 0657-56 &$10.6\pm 0.2$ & 2 & $18\pm 3$ & 2 & $191\pm32$ \\
   & $10.7\pm0.4$& 1& $17.6\pm0.2$ & 1 & $188.3\pm19.5$\\
   & $11.70\pm0.22$&3&$26.1 \pm 2.4$&3&$305.4\pm28.7$\\
Abell 907 &$5.8\pm 0.3$ & 1 & $4.3\pm 0.5$ & 1 & $24.9\pm3.2$ \\
XMMXCSJ095940.8+023111.3 &$7.3_{-0.8}^{+1.4}$ &9 & \nodata &\nodata &\nodata\\ 
RXCJ1023.6+0411 &$8.38\pm 0.44$ & 3 & $10.8\pm 1.0$ & 3 & $90.5\pm9.6$ \\
MS1054.4-0321 &$8.3\pm 0.7$ & 14 & \nodata &\nodata &\nodata \\
MACSJ1115.8+0129 &$9.20\pm 0.98$ & 3 & $9.9\pm 1.1$ & 3 & $91.1\pm14.0$\\
Abell 1300 &$9.2\pm 0.4$ & 2 & $8\pm 2$ & 2 & $74\pm19$ \\
  &$9.35\pm1.67$ &3&$21.6 \pm2.8$ & 3&$202.0\pm44.6$\\
RXCJ1206.2-0848 &$10.71\pm 1.29$ & 3 & $21.9\pm 2.9$ & 3 & $234.5\pm42.0$\\ 
XMMUJ1230.3+1339 &$6.0_{-1.2}^{+1.6}$ &4 & $3\pm 1$ & 4 & $18_{-6}^{+7}$ \\
RDCSJ1252-2927 &$6.6_{-1.2}^{+1.5}$ &6 & $0.66_{-0.10}^{+0.09}$ & 6 & $4.4_{-1.0}^{+1.2}$\\
MACSJ1311.0-0311 &$6.2\pm 0.7$ & 11 & $4.6_{-0.1}^{+0.1}$ & 11 & $28.5_{-3.3}^{+3.3}$\\
Abell 1689 &$8.5\pm 0.2$ & 1 & $10.5\pm 1.4$ & 1 & $89.2\pm12.1$ \\
RXCJ1347.5-1144 &$10.75\pm 0.83$ & 3 & $24.8\pm 2.7$ & 3 & $266.6\pm35.6$\\
MACSJ1359.1-1929 &$6.73\pm 0.96$ & 13 & \nodata &\nodata &\nodata \\
Abell 1835 &$8.4\pm 0.3$ & 1 & $11.8\pm 1.4$ & 1 & $99.1\pm12.3$ \\
   &$9.00\pm0.25$&3&$14.1\pm0.6$&3&$126.9\pm 6.4$\\
RXCJ1504.1-0248 &$8.00\pm 0.44$ & 3 & $12.5\pm 1.3$ & 3 & $100.0\pm11.8$\\ 
Abell 2163 &$12.27\pm 0.90$ & 3 & $44.0\pm 4.5$ & 3 & $539.9\pm67.9$ \\
Abell 2204 &$7.6\pm 0.2$ & 1 & $8.4\pm 1.0$ & 1 & $63.8\pm7.8$ \\
MACSJ1931.8-2635 &$7.47\pm 1.40$ & 3 & $11.4\pm 1.5$ & 3 & $85.2\pm19.5$ \\
RXCJ2011.3-5725 &$3.23 \pm 0.34$ & 3 & $3.7 \pm0.7$& 3& $12.0\pm 2.6$\\
RXCJ2014.8-2430 &$5.63\pm 0.11$ & 12 & $7.1_{-0.1}^{+0.1}$ & 12 & $39.9_{-1.1}^{+1.1}$\\ 
MACSJ2046.0-3430 &$5.81\pm 1.02$ & 13 & \nodata &\nodata &\nodata \\
RXCJ2214.9-1359 &$8.8\pm 0.7$ & 8 & \nodata &\nodata &\nodata \\
XMMXCSJ2215.9-1738 &$7.4_{-1.4}^{+2.1}$ &6 & $0.38\pm 0.09$ & 6 & $2.8_{-0.9}^{+1.0}$\\ 
XMMUJ2235.3-2557 &$6.0_{-1.8}^{+2.5}$ &5 & $0.95_{-0.12}^{+0.11}$ & 6 & $5.7_{-1.9}^{+2.5}$ \\
\hline
\end{tabular}
\end{center}
\end{minipage}
\end{table*}

\begin{table*}
\begin{minipage}{105mm}
\begin{center}
\contcaption{Summary of cluster X-ray properties.}
\begin{tabular}{|l |c|c|c|c|c|}
\hline\vspace{-3.5ex}\\
Cluster Name &$T_X$ &Ref. & $M_{\rm{gas}}$ &  Ref.&  $Y_X$  \\
  & keV &   & $10^{13} \rm{M}_{\odot}$&  & $10^{13}$keV\,$\rm{M}_{\odot}$\\
\hline\vspace{-2.5ex}\\
RXCJ2243.3-0935 &$8.24\pm 0.92$ & 3 & $19.8\pm 2.4$ & 3 & $163.2\pm26.9$ \\
Abell S1077 &$8.0\pm 0.5$ & 10 & \nodata &\nodata &\nodata \\
Abell 2537 &$7.9\pm 0.7$ & 2 & $7\pm 1$ & 2 & $55\pm9$ \\
   &$7.6\pm0.7$ & 1& $6.5\pm 0.7$&1& $49.4\pm7.0$\\
 &$7.63\pm0.86$&3&$8.2\pm1.1$&3&$62.6\pm11.0$\\
RXCJ2337.6+0016 &$9.6\pm 0.3$ & 2 & $8\pm 1$ & 2 & $77\pm10$ \\
    &$7.5\pm0.5$ & 1&$8.1\pm 0.7$&1&$60.8\pm6.6$\\
\hline
\end{tabular}
\begin{enumerate}
\renewcommand{\theenumi}{(\arabic{enumi})}
\item \citet{zhang2008} $kT_{(0.2-0.5)r_{500}}$  
\item \citet{zhang2006} $kT_{(0.1-0.5)r_{500} }$
\item \citet{mantz2010b} $kT_{(0.15-0.5)r_{500}}$
\item \citet{fassbender2011} $kT_{r<71\arcsec}$
\item \citet{mullis2005} $kT_{r<50\arcsec}$
\item \citet{culverhouse2010} $kT_{r<30\arcsec}$
\item \citet{pacaud2007} $kT_{r<80\arcsec}$
\item \citet{ebeling2007} $kT_{70kpc <r < r_{1000}}$
\item \citet{mehrtens2012}
\item \citet{defilippis2004} $kT_{r<2.5\arcmin}$
 \item \citet{maughan2008} $kT_{(0.15-1)r_{500}}$
 \item \citet{pratt2009} $kT_{(0.15-1)r_{500}}$
\item \citet{allen2008} $kT_{r < r_{2500}}$
 \item \citet{branchesi2007} $kT_{r<84\arcsec}$
\end{enumerate}
\end{center}
\end{minipage}
\end{table*}


\section{APEX-SZ Data Analysis}
\label{SEC:szdatareduction}
Raw APEX-SZ timestream data contain the faint SZE cluster signal, scan synchronous signals due to ground pick-up and instrumental thermal fluctuations, atmospheric fluctuations, and instrument noise.   In order to measure the SZE surface brightness accurately, the array properties must be well characterized and contamination cleaned from the data.  

\subsection{Beams and Calibration}
The beam position and shape for each bolometer in the focal plane are measured from a daily raster scan of a calibration target (Mars, Uranus, or Saturn\footnote{Saturn is used solely to measure beam profiles. It is not used for absolute flux calibration as its large signal can saturate the APEX-SZ bolometers.}).   In addition to the 58 arcsec FWHM Gaussian main beam, APEX-SZ beams exhibit significant sidelobes \citep{schwan2011}.  We characterize the sidelobes by mapping the individual detector beams and combining them into a composite beam.  The total beam area, including the Gaussian main beam and the sidelobes, is then taken into account during further calibration.  The beam size is also corrected to account for the angular extent of the source and detector saturation effects.  

Absolute flux calibration is performed based on the overall amplitude of response for each detector from a raster scan on a known celestial source.  APEX-SZ observed two primary flux calibrators: Mars and Uranus. The planetary disk size and brightness temperature are taken from the Rudy model for Mars \citep{rudy1987, muhleman1991} and from the JCMT FLUXES\footnote{http://www.jach.hawaii.edu/jac-bin/planetflux.pl} model for Uranus.  As described in \citet{halverson2009}, we refine the absolute brightness temperatures for Mars and Uranus using \textit{WMAP} results \citep{hinshaw2009, weiland2011} to improve calibration accuracy.  We find that the Rudy temperatures are systematically higher than \textit{WMAP} by a factor of $1.052 \pm 0.01$.  The calculation of this factor includes the extrapolation of the Rudy and \textit{WMAP} 94 GHz brightness temperatures to 150 GHz, 1.016$\pm$ 0.009, which is included in the total calibration uncertainty.   Similarly, we use the data of \citet{griffin1993} to extrapolate the \textit{WMAP7} 94 GHz measurement of Uranus to $T_{150 \rm{GHz}}  = 100.4 \pm 5.3$ K.

Several times during observations primary planetary calibrators were unavailable due to angular proximity to the Sun.   The stable Galactic H II regions RCW38 and IRAS12073-6233 were used as secondary calibrators during these periods, with brightness temperatures bootstrapped from back-to-back scans of the primary and secondary calibrators.

The measured calibration is corrected for differences in atmospheric opacity between the data and calibration scans using the measured value of precipitable water vapour (PWV) from the APEX radiometer.  Following \citet{sayers2011}, we extrapolate the atmospheric optical depth based on the ATM model \citep{pardo2001} and correct accordingly. 

In addition to the calibration scan, routine pointing scans of bright quasars are performed throughout the night. We find an average pointing jitter of approximately 5.7 arcsec in both azimuth and elevation.  This is significantly less than the size of the APEX-SZ beam and has a negligible effect on the measurements presented here.

Combining the corrections discussed in this section (beam solid angle, calibrator temperature, atmospheric opacity) with the uncertainty in the frequency band center \citep[see][]{halverson2009} we estimate the Gaussian uncertainty in the flux calibration to be $\pm$8 per cent. 

\subsubsection{Gain Fluctuations}
Ideally, the responsivity of each bolometer is constant despite changing amounts of incident optical power \citep{lee1996}. However, when a bolometer is biased near the upper edge of the superconducting transition, the responsivity changes as a result of varying optical power from three sources: the elevation dependence of atmospheric airmass, the overall atmospheric opacity, and the power from an astronomical source. The last effect is sub-dominant, but it cannot be neglected for the case of a bright calibrator.  For most detectors in the APEX-SZ array, suppression of responsivity is approximately a linear function of incident optical power. However, some channels exhibit more extreme changes.

We measure the responsivity of each bolometer using the elevation($\epsilon$) dependent signal $d_i \propto A_{\csc,i} \csc(\epsilon)$ introduced by the circular scan pattern.  This is repeated for every two minutes of data, normalizing by the changing zenith emissivity to measure $A_{\csc,i}$.   Channels that exhibit strong nonlinear response are discarded from use in further analysis.   For the remaining channels, we model the response as a linear function and use it  to correct the absolute flux calibration, discarding data that deviate more than $2.5\sigma$ from the best-fit model.   Data with extremely large overall correction factors (greater than 50 per cent) are discarded.  

We also correct for two second order effects.  First, as a bright calibration source is scanned, variable responsivity will slightly distort the measured beam shape. Second, the fractional solid angle of the sidelobes relative to the main beam will be affected. 

Based on the rms scatter around the best-fit gain models, we estimate the uncertainty in the responsivity corrections to be $\pm$5 per cent.  Combining this uncertainty with the calibration uncertainty from the previous section, we find that the total uncertainty in the APEX-SZ temperature calibration is $\pm$10 per cent.

\subsection{Timestream Processing}

The APEX-SZ data set was measured in widely varying atmospheric conditions, a significant challenge when trying to measure the relatively faint SZE signal.  In general, the atmospheric signal is highly correlated between bolometers as spatial fluctuations in the atmosphere are much larger than the APEX-SZ FOV.  Additionally, the bolometers exhibit a correlated signal due to heating of the detector array from telescope vibrations.  We employ several different filtering techniques to clean the data, similar to \citet{halverson2009}.  In order to optimize  the final maps, filtering is tailored to the specific needs of each target with the goal of sufficiently removing contamination and minimizing noise while preserving cluster signal.  

\subsubsection{Initial Data Conditioning}
The first step to condition APEX-SZ timestreams is to parse the data into separate circles made on the sky according to the scan pattern.  The circles are grouped together based on a common central azimuth and elevation and data not in a circle set are discarded.

Initial channel and data cuts are performed on the indexed data based on several criteria. Channels that are optically unresponsive (optical efficiency less than 6 per cent) are rejected.     
APEX-SZ timestreams occasionally exhibit large spikes or glitches in one or many bolometers due to either cosmic rays or electrical interference.  Glitches are located and removed from the timestreams using both the timestream derivative and a simple S/N threshold cut.  When a glitch is simultaneously detected in several channels (more than 7 per cent), that particular section of data is discarded for all channels.  As a final step, channels with excess noise between 12 and 18 Hz are discarded.

After the initial data cuts we deconvolve an optical time constant for each channel from the timestreams.  The median measured optical time constant for APEX-SZ bolometers is $\sim$12 ms, however, it is as high as $\sim$60 ms for some channels.  To prevent the amplification of high frequency noise during deconvolution, an additional low-pass filter ($F = \exp\left[ -(f/f_0)^6\right]$) is applied.  We define $f_0=2*v_{\rm{scan}}/\rm{FWHM}$ where $v_{\rm{scan}}$ is the median scan velocity of the telescope and $\rm{FWHM}$ is the full width at half-maximum of the APEX-SZ beam.

\subsubsection{Removal of Timestream Contamination}

The final step in processing the timestream data is to remove the dominant atmospheric and thermal signals.   The majority of the timestream cleaning is performed through a combination of two methods: high-pass time-domain filters and the subtraction of spatial polynomial functions across the array, taking advantage of the correlation between bolometers.

An initial high-pass filter removes the static offset for each channel as well as low-frequency noise in the timestreams.  Removing the timestream signal due to stage heating is especially important because the subsequent spatial template removal steps use relative gain coefficients that are based solely on the atmospheric response.  
The high-pass filtering is implemented in two forms: an N-order polynomial removal and a Butterworth frequency domain filter.  The polynomial filter, described in detail in \citet{halverson2009}, simultaneously removes a scan synchronous cosecant signal.
\begin{figure}\centering
\includegraphics[width=0.52\textwidth]{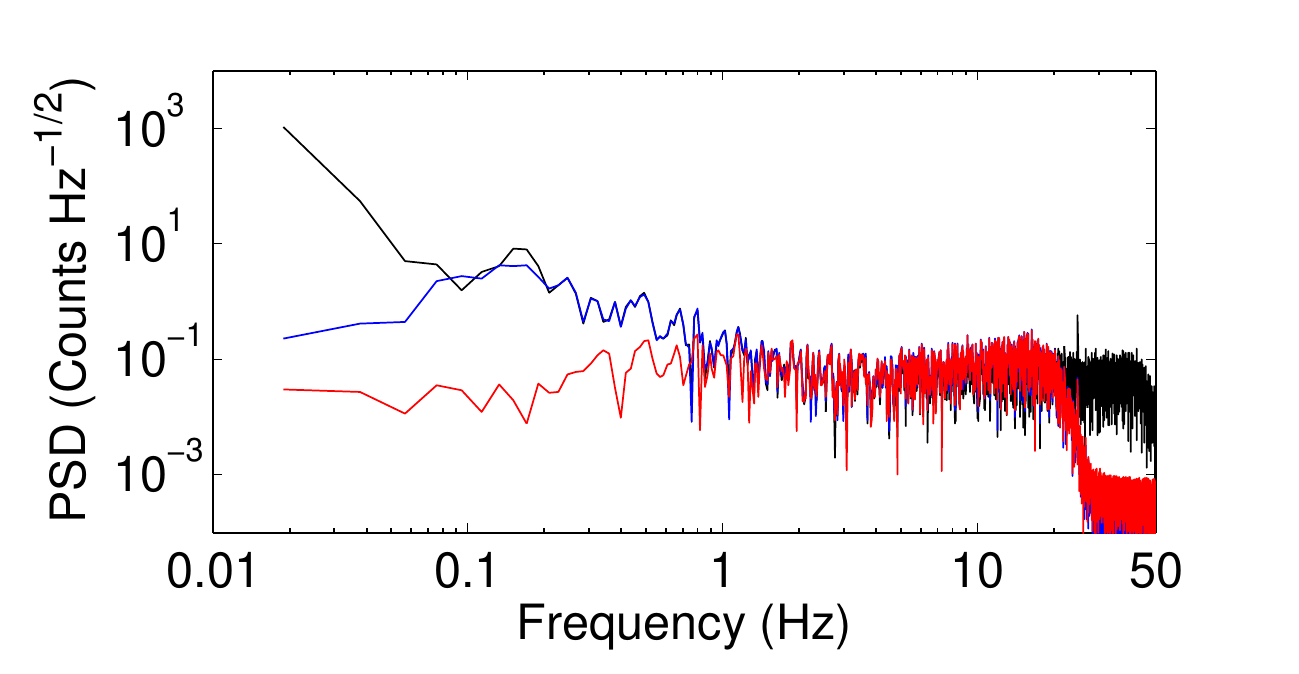}
\caption[]{The power spectral density (PSD) of an APEX-SZ timestream at three different stages of timestream reduction.  The black line represents the raw data in units of readout counts.  The blue line shows the power remaining after initial data cuts are performed, optical time constants are deconvolved, and a high-pass filter is applied.  The decrease in low-frequency power is due to the high-pass step, whereas roll-off at high-frequencies is a result of the low-pass filter included in the time constant deconvolution.  The red line shows the power spectrum after spatial template removal and polynomial removal steps are performed, removing the remaining low-frequency contamination and completing the cleaning of the data.}
\label{fig:tspowspect}
\end{figure}

After the first high-pass filter, we remove a 2-dimensional spatial polynomial function across the array \citep{sayers07,halverson2009} at each sample in the timestream.  During this process, referred to as \textit{spatial template removal}, we fit the polynomial either across the entire array, or individually for each of the six detector wafers to improve removal of correlated signals on scales smaller than the $0.4^\circ$ field-of-view.  Often, this filter is applied multiple times, discarding channels with excess variance in between.  A final polynomial removal is performed to remove any remaining contamination.  Figure \ref{fig:tspowspect} shows the power spectrum of an APEX-SZ timestream at three different steps in the reduction process.

\subsection{Map Making \& Transfer Function}
\label{SEC:transferfunc}
\begin{figure*}\centering
\includegraphics[width=0.9\textwidth]{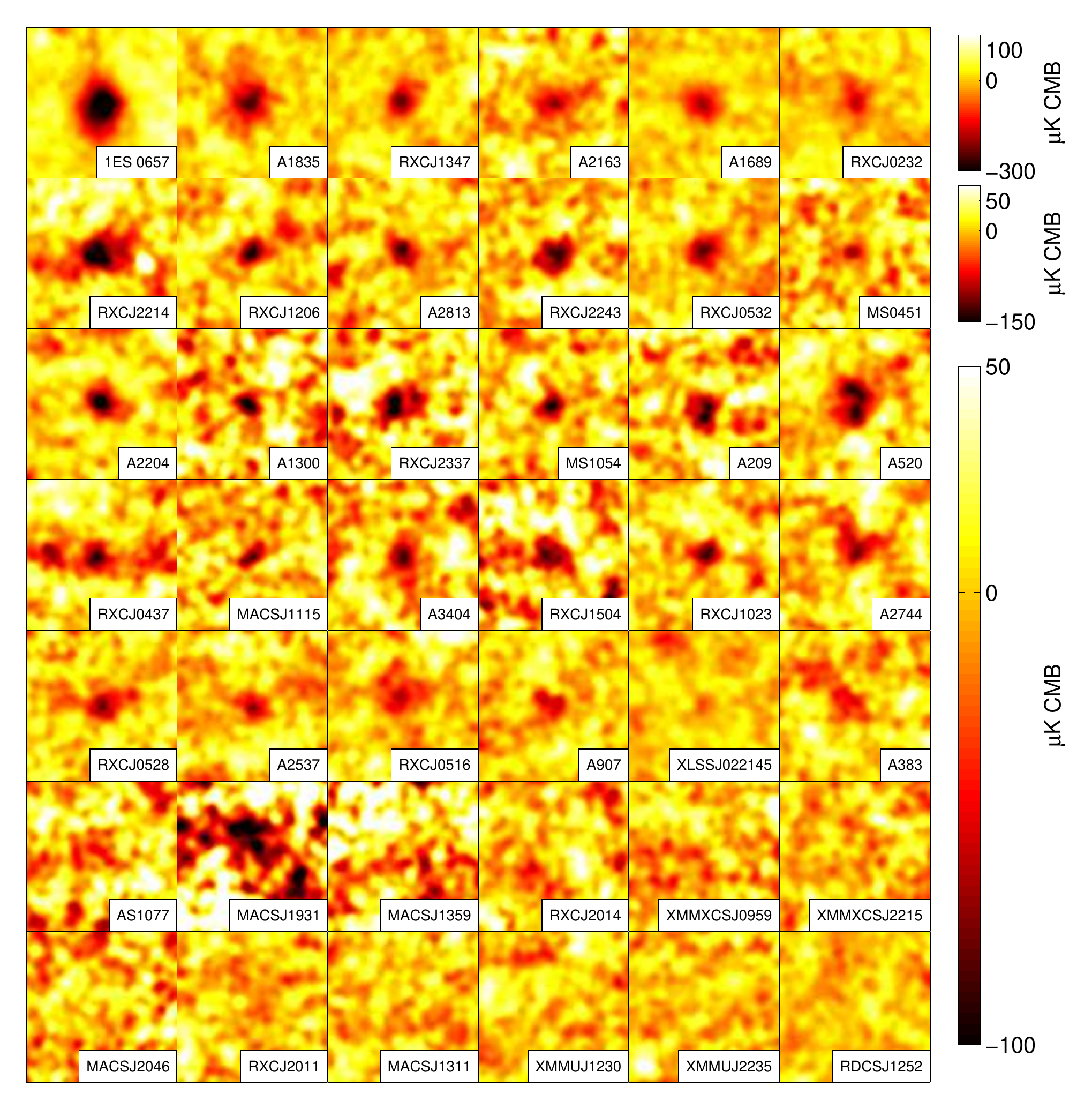}
\caption[]{Sky maps for each of the APEX-SZ cluster positions in signal-to-noise units.  The data for each 15 x 15 square arcmin map have been filtered to remove contaminating signals.  Therefore, the maps show an attenuated representation of the sky.  Each map  has been additionally convolved with a one arcmin Gaussian.   
}
\label{fig:skymaps}
\end{figure*}

Once the timestreams are filtered, we create sky maps by binning the timestream data into $10 \times 10$ arcsec pixels. Sky coordinates for each bolometer are determined from the absolute telescope boresight pointing and individual bolometer positional offsets.  As in \citet{halverson2009}, bolometers and scans are combined with a minimum variance weighting for each cluster.
The resulting set of 42 sky maps is shown in Figure \ref{fig:skymaps}.  Each map has been convolved with a one arcmin FWHM Gaussian for visual presentation. However, all subsequent analysis is performed on the unsmoothed version.

In conjunction with the sky map, we quantify the effects of the APEX-SZ instrumental beam and analysis filtering on the sky signal.  A simulated cluster map, created using the profile discussed in Section \ref{SEC:szmmodel},  is convolved with the composite beam profile of the array and mapped into timestreams using the telescope pointing information.  Each data cut and filtering process performed on the real cluster data is repeated on these timestreams.  The filtered simulation is mapped and coadded using the same minimum variance weights as the cluster map.  An individualized transfer function map, $\mathbf{K}$,  is created for each cluster by normalizing the Fourier transform of the filtered map to that of the original simulation.

We create difference noise maps for each cluster to characterize the map noise. In this process, random halves of the individual scan maps are multiplied by -1 before coadding.
This removes any astrophysical signal that is consistent between all maps, including primary CMB anisotropy.  To account for the noise due to the CMB, we convolve a realization of the CMB created from the Planck + WP best-fit power spectrum \citep{2014planckxvi} with a CMB transfer function (created in the same way as the cluster transfer function) and add it to the difference noise map.  In addition to the CMB, a contribution from a background of point sources is expected.  We find that the power from Poisson distributed point sources \citep{hall2010,shirokoff2010}, including an extra 50 per cent variance due to lensing \citep{hezaveh2013},  is negligible in comparison to the instrument noise.

\subsection{Cluster Signal Modeling}
\label{SEC:clustermodel}

\subsubsection{Parametric Models}
\label{SEC:szmmodel}
In order to calculate the integrated cluster signal, we need to extrapolate the measured cluster signal to spatial scales that have been filtered out due to timestream processing and the angular resolution of the instrument.   We adopt the universal pressure profile of \citet{arnaud2010},

\begin{equation}
\label{EQN:gnfwmodel}
P_e(r)=\frac{P_{0}}{(r/r_s)^\gamma[1+(r/r_s)^\alpha]^{(\beta-\gamma)/\alpha}}.
\end{equation}
Here $P_0$ is the normalization of the pressure profile, $r_s$ is the scale radius, and ($\alpha, \beta, \gamma$) describe the slope of the profile for $r \approx r_s$, $r>r_s$ and $r<r_s$, respectively.  The scale radius is often defined in terms of a concentration parameter $c_{500}$, $r_s = r_{500}/c_{500}$.  Proposed by \citet{nagai2007a}, this generalized Navarro-Frenk-White model (GNFW), is a generalization of the numerical results from \citet{navarro1997} for the distribution of mass in a dark matter halo.   \citet{arnaud2010} combined simulations and X-ray observations to measure a best-fit model of ($c_{500}$, $\alpha, \beta, \gamma$) = (1.177, 1.0510, 5.4905, 0.3081), which we use for the measurements presented here.  Applying the ideal gas law, $P_e = n_ekT_e$, to Equation \ref{EQN:gnfwmodel} and substituting into Equation \ref{EQN:szeffect} the SZE profile becomes
\begin{eqnarray}
\label{EQN:szgnfwmodel}
\frac{\Delta T_{\rm{SZE}}}{T_{\rm{CMB}}}& = & \int f(x,T_e)\cdot \frac{\sigma_T}{m_ec^2} P_e(r)dl.
\end{eqnarray}
We reparametrize  Equation \ref{EQN:szgnfwmodel} in terms of a normalization $\Delta T_0$,
\begin{equation}
\Delta T_{\rm{SZE}} = \Delta T_0 \int \frac{1}{(r/r_s)^\gamma[1+(r/r_s)^\alpha]^{(\beta-\gamma)/\alpha}}dl.
\end{equation}
Due to the compact scan pattern and strong atmospheric filtering required, the APEX-SZ data poorly constrain power on scales larger than 10 arcmin in the maps.  Therefore, we infer $r_{500}$ from the X-ray based $r_{500}-T$ scaling relation of \citet{vikhlinin2006} for each cluster and fix the model centroid at the X-ray centroid.  The \citet{arnaud2010} profile for each cluster serves as the simulated input for creating the individualized transfer functions described in $\S$\ref{SEC:transferfunc} and is used to fit for the normalization, $\Delta T_0$.

\subsubsection{Likelihood Analysis}

We calculate the likelihood for each model as a function of $\Delta T_0$, $\mathcal{L} \propto e^{-\chi^2(\Delta T_0)/2}$.  
First, we create a 2-dimensional model map $\mathbf{M}(\Delta T_0)$ and convolve it with the transfer function map $\mathbf{K}$, $\mathbf{M}'(\Delta T_0)= \mathbf{K} * \mathbf{M}(\Delta T_0)$.   The model $\mathbf{M}'(\Delta T_0)$ and sky maps $\mathbf{D}$ are binned into 1-arcmin radial bins around the X-ray centroid.  Using the central 10 arcmin of these profiles, we construct the $\chi^2$ statistic,
\begin{equation}
\label{EQN:chisq}
\chi^{2} = (\mathbf{D} - \mathbf{M}'(\Delta T_0))^{\rm{T}}\mathbf{C}_{n}^{-1}(\mathbf{D} - \mathbf{M}'(\Delta T_0)),
\end{equation}
where $\mathbf{C}_n$ is the noise covariance matrix estimated by averaging 500 radially-binned difference noise realizations.  The best-fit parameters along with the X-ray based $\theta_{500} = r_{500}/d_A$ and respective probability to exceed (PTE) are given in Table \ref{TBL:mcmc}.   

For each cluster, we calculate the spherical integral of the assumed model profile within the X-ray defined $r_{500}$ and use the measured $\Delta T_0$ to estimate the integrated Comptonization $Y_{500}$.  We use the work of \citet{nozawa2000} and the X-ray temperatures from Table \ref{TBL:physparam} to calculate the relativistic correction $\delta_{\rm{SZE}}$ to the frequency dependent term in Equation \ref{EQN:szefreq}.   On average, this correction is of order $\sim5$ per cent.  The measured values of $Y_{500}$ are given in Table \ref{TBL:mcmc}.


\section{Results}
\label{SEC:results}

\subsection{Cluster Observables}

We use the method described in Section \ref{SEC:clustermodel} to measure the spherically-integrated Comptonization for each cluster.  We assume the cluster follows an \citet{arnaud2010} pressure profile (see Equation \ref{EQN:gnfwmodel})  with radius $r_{500}$, where $r_{500}$ is inferred from the \citet{vikhlinin2006} $r-T_X$ relation.

We use three different proxies for total cluster mass: spectroscopic X-ray temperature, gas mass, and $Y_X$, the values for which are taken from the literature. 
The redshifts and X-ray parameters for the 42 clusters included in this sample are given in Table \ref{TBL:physparam} along with the references.  For clusters found in more than one of the REFLEX-DXL, Z08, or M10 subsamples, the additional parameter values are also given.  All measurements were made with either the \textit{XMM-Newton} or \textit{Chandra} X-ray observatories.  X-ray temperatures are used that were derived from temperature maps with the central region excluded when available.  Inclusion of the core region can lead to an underestimated global temperature for the cool-core clusters present within our sample. $M_{\rm{gas}}$ values, and by extension $Y_X$, were available for only 35 clusters.

\begin{table*}
\begin{minipage}{126mm}
\begin{center}

\caption{APEX-SZ Model Fits}
\label{TBL:mcmc}
\begin{tabular}{lccccc}
\hline\vspace{-3.5ex}\\
Cluster Name &$\Delta T_0$ & $\theta_{500}$ &Ref. &PTE &$Y_{500}$\\ 
    &($\mu$K CMB)& ($\arcsec$)&  &  &($10^{-10} sr$) \\
\hline\vspace{-2.5ex}\\
Abell 2744 &   $-439\pm80$ &$311.3$&2& 0.11 & $1.51\pm0.28$ \\
  &$-433\pm80$ &$284.8$&3& 0.07 & $1.23\pm0.23$ \\
Abell 2813 &   $-482\pm74$ &$281.7$&2& 0.14 & $1.33\pm0.20$ \\
  &$-498\pm78$ &$267.9$&1& 0.24 & $1.24\pm0.19$ \\
  &$-503\pm80$ &$279.6$&3& 0.14 & $1.37\pm0.22$ \\
Abell 209 &   $-472\pm76$ &$360.9$&1& 0.24 & $2.14\pm0.34$ \\
  &$-440\pm78$ &$390.1$&3& 0.32 & $2.34\pm0.42$ \\
XLSS J022145.2-034617 &   $-200\pm48$ &$158.6$&8& 0.21 & $0.17\pm0.04$ \\
RXCJ0232.2-4420 &   $-611\pm73$ &$274.3$&2& $7.8\times10^{-3}$ & $1.60\pm0.19$ \\
  &$-622\pm67$ &$266.0$&1& $4.1\times10^{-3}$ & $1.52\pm0.16$ \\
  &$-581\pm68$ &$332.1$&3& $7.8\times10^{-3}$ & $2.27\pm0.26$ \\
Abell 383 &   $-264\pm71$ &$320.5$&1& 0.79 & $0.93\pm0.25$ \\
RXCJ0437.1+0043 &   $-341\pm72$ &$232.2$&2& $3.8\times10^{-5}$ & $0.63\pm0.13$ \\
MS0451.6-0305 &   $-638\pm159$ &$153.2$&12& 0.21 & $0.52\pm0.13$ \\
Abell 520 &   $-354\pm50$ &$373.9$&4& $1.2\times10^{-5}$ & $1.72\pm0.24$ \\
RXCJ0516.6-5430 &   $-221\pm54$ &$276.2$&2& $1.3\times10^{-8}$ & $0.59\pm0.14$ \\
  &$-221\pm59$ &$260.3$&1& $3.6\times10^{-9}$ & $0.52\pm0.14$ \\
RXCJ0528.9-3927 &   $-418\pm64$ &$278.5$&2& $1.2\times10^{-4}$ & $1.13\pm0.17$ \\
  &$-426\pm61$ &$266.0$&1& $5.5\times10^{-5}$ & $1.04\pm0.15$ \\
  &$-444\pm61$ &$290.4$&3& $1.9\times10^{-5}$ & $1.31\pm0.18$ \\
RXCJ0532.9-3701 &   $-572\pm79$ &$331.3$&2& $5.6\times10^{-3}$ & $2.21\pm0.31$ \\
  &$-553\pm76$ &$296.6$&1& $4.5\times10^{-3}$ & $1.70\pm0.23$ \\
Abell 3404 &   $-462\pm64$ &$464.4$&1& $8.4\times10^{-3}$ & $3.47\pm0.48$ \\
1ES 0657-56 &   $-1113\pm49$ &$329.5$&2& $3.3\times10^{-6}$ & $4.27\pm0.19$ \\
  &$-1123\pm47$ &$331.1$&1& $3.6\times10^{-6}$ & $4.35\pm0.18$ \\
  &$-1106\pm49$ &$347.1$&3& $8.5\times10^{-6}$ & $4.74\pm0.21$ \\
Abell 907 &   $-334\pm52$ &$411.8$&1& 0.71 & $1.95\pm0.30$ \\
XMMXCSJ095940.8+023111.3 &   $-112\pm93$ &$139.2$&10& $1.4\times10^{-3}$ & $0.076\pm0.063$ \\
RXCJ1023.6+0411 &   $-440\pm62$ &$295.7$&4& 0.16 & $1.35\pm0.19$ \\
MS1054.4-0321 &   $-808\pm106$ &$123.8$&15& 0.98 & $0.43\pm0.06$ \\
MACSJ1115.8+0129 &   $-290\pm88$ &$262.1$&4& 0.12 & $0.70\pm0.21$ \\
Abell 1300 &   $-497\pm97$ &$295.6$&2& 0.73 & $1.53\pm0.30$ \\
  &$-458\pm96$ &$298.1$&3& 0.83 & $1.43\pm0.30$ \\
RXCJ1206.2-0848 &   $-756\pm103$ &$237.8$&4& 0.18 & $1.51\pm0.21$ \\
XMMUJ1230.3+1339 &   $-204\pm97$ &$91.6$&5& 0.31 & $0.059\pm0.028$ \\
RDCSJ1252-2927$^*$ &   $-22\pm90$ &$79.2$&7& 0.28 & $(4.89\times10^{-3}) \pm 0.020$\\
MACSJ1311.0-0311$^*$ &   $-72\pm85$ &$161.8$&12& 0.88 & $0.066\pm0.077$ \\
Abell 1689 &   $-1043\pm93$ &$444.3$&1& 0.23 & $7.21\pm0.64$ \\
RXCJ1347.5-1144 &   $-1002\pm126$ &$233.1$&4& 0.24 & $1.93\pm0.24$ \\
MACSJ1359.1-1929$^*$ &   $-109\pm112$ &$183.5$&14& 0.02 & $0.13\pm0.13$ \\
Abell 1835 &   $-810\pm105$ &$333.7$&1& 0.42 & $3.16\pm0.41$ \\
  &$-770\pm118$ &$346.0$&3& 0.36 & $3.24\pm0.50$ \\
RXCJ1504.1-0248 &   $-584\pm112$ &$374.4$&4& 0.52 & $2.86\pm0.55$ \\
Abell 2163 &   $-684\pm105$ &$494.0$&4& 0.90 & $5.96\pm0.92$ \\
Abell 2204 &   $-623\pm52$ &$497.4$&4& 0.01 & $5.37\pm0.45$ \\
MACSJ1931.8-2635$^*$ &   $-37\pm117$ &$236.6$&4& 0.20 & $0.072\pm0.228$ \\
RXCJ2011.3-5725   &$-109\pm78$ &$185.6$&3& 0.16 & $0.13\pm0.09$ \\
RXCJ2014.8-2430$^*$ &   $-26\pm89$ &$403.1$&13& 0.58 & $0.14\pm0.50$ \\
MACSJ2046.0-3430 &   $-132\pm108$ &$177.7$&14& 0.72 & $0.14\pm0.12$ \\
RXCJ2214.9-1359 &   $-730\pm91$ &$191.8$&9& $4.4\times10^{-4}$ & $0.94\pm0.12$ \\
XMMXCSJ2215.9-1738$^*$ &   $-14\pm145$ &$74.1$&7& 0.24 & $(2.68\times10^{-3}) \pm 0.028$\\
XMMUJ2235.3-2557$^*$ &   $33\pm153$ &$68.6$&6& 0.51 & $(-5.35\times10^{-3}) \pm 0.025$\\
RXCJ2243.3-0935 &   $-846\pm111$ &$204.1$&4& 0.15 & $1.23\pm0.16$ \\
Abell S1077 &   $-234\pm112$ &$270.9$&11& 0.08 & $0.60\pm0.29$ \\
Abell 2537 &   $-408\pm70$ &$281.7$&2& 0.02 & $1.13\pm0.19$ \\
  &$-424\pm71$ &$276.0$&1& 0.03 & $1.13\pm0.19$ \\
  &$-398\pm75$ &$276.6$&3& 0.02 & $1.06\pm0.20$ \\
RXCJ2337.6+0016 &   $-398\pm73$ &$330.1$&2& 0.12 & $1.53\pm0.28$ \\
  &$-386\pm79$ &$289.9$&1& 0.18 & $1.13\pm0.23$ \\
\hline\vspace{-3.5ex}\\
\end{tabular}
\end{center}
$Y_{500}$ is the spherical integration of best-fit
  pressure profile within $r_{500}$.  Non-detections are denoted with
  $^{*}$.  References correspond to those in Table \ref{TBL:physparam}
  and denote the X-ray measurement of $kT$  used to determine
  $r_{500}$.  Values are given for each subsample where appropriate:
  [1] Z08, [2] REFLEX-DXL, [3] M10.

\end{minipage}
\end{table*}

\subsection{Regression Analysis}
\begin{figure*}
  \centering
\begin{tabular}{lcr}
\includegraphics[width=0.32\textwidth]{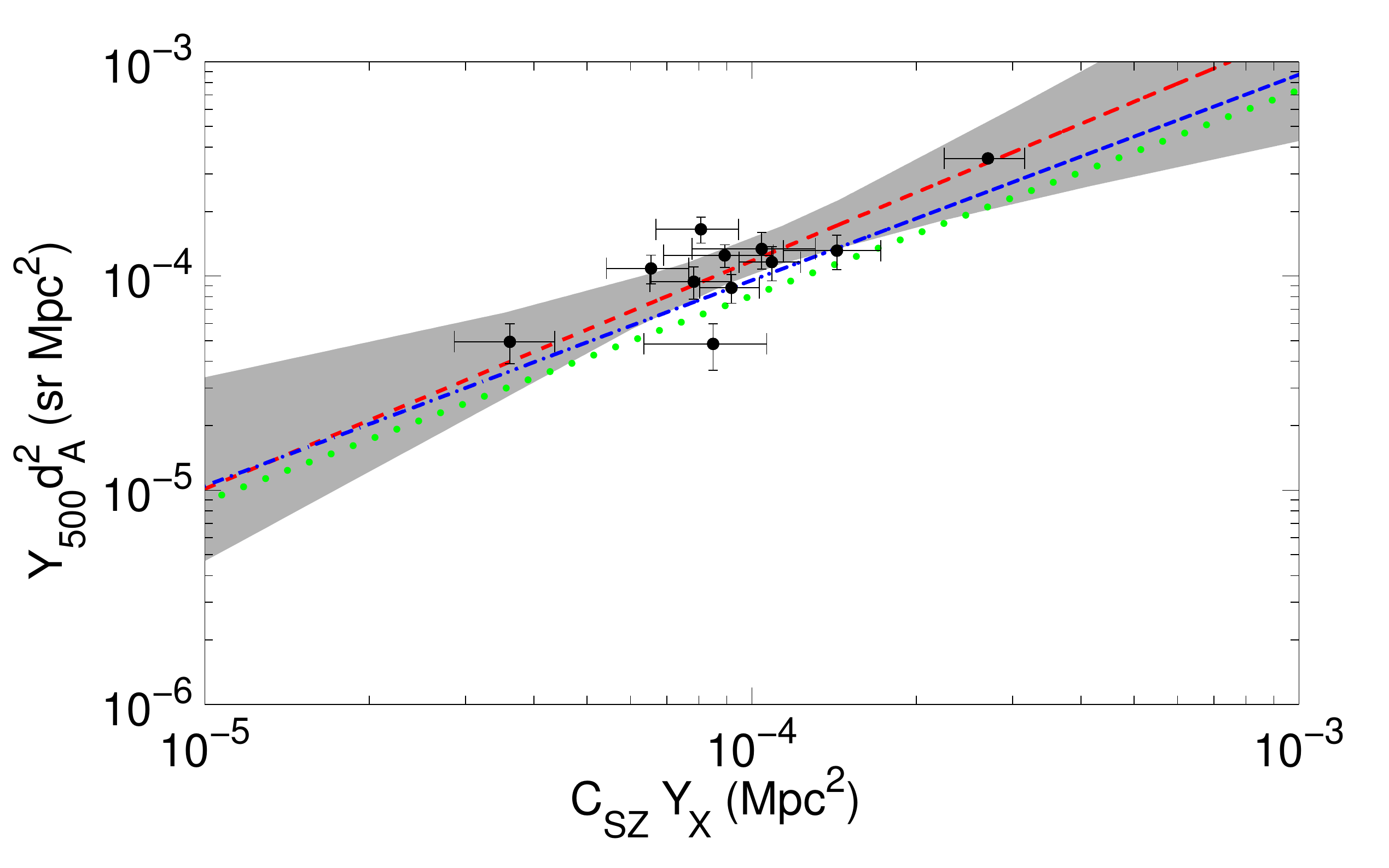}&
\includegraphics[width=0.32\textwidth]{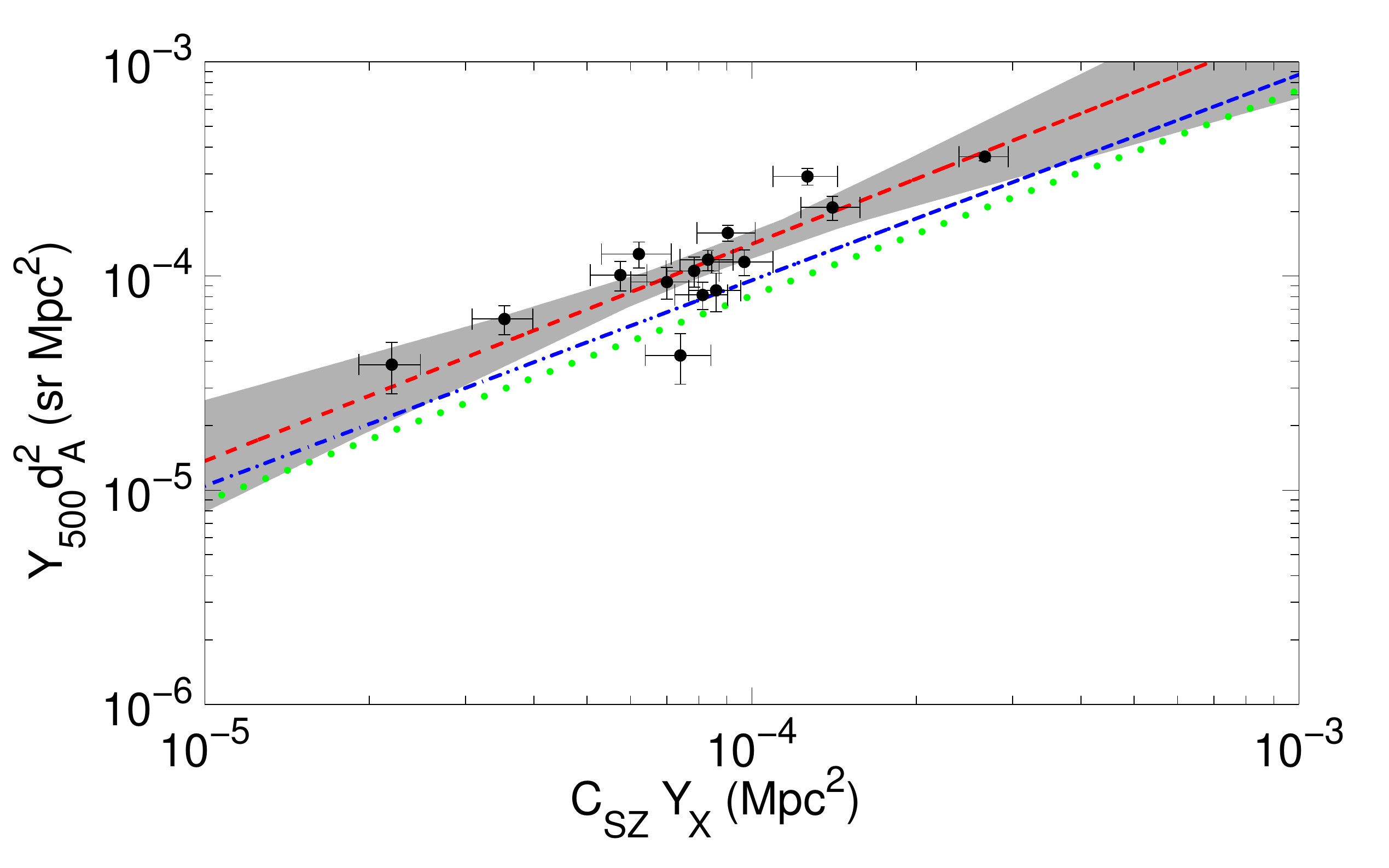}&
\includegraphics[width=0.32\textwidth]{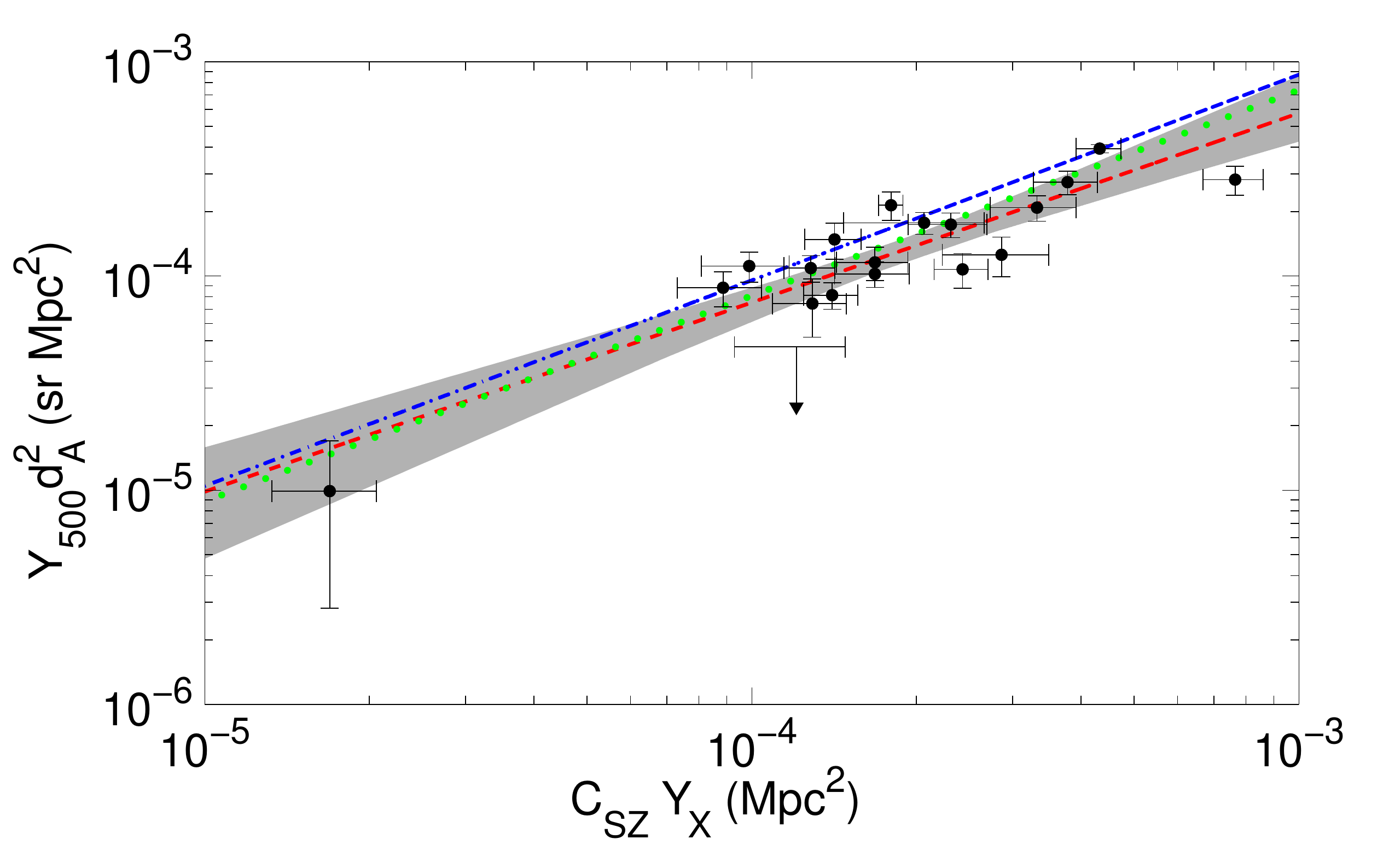}\\
\includegraphics[width=0.32\textwidth]{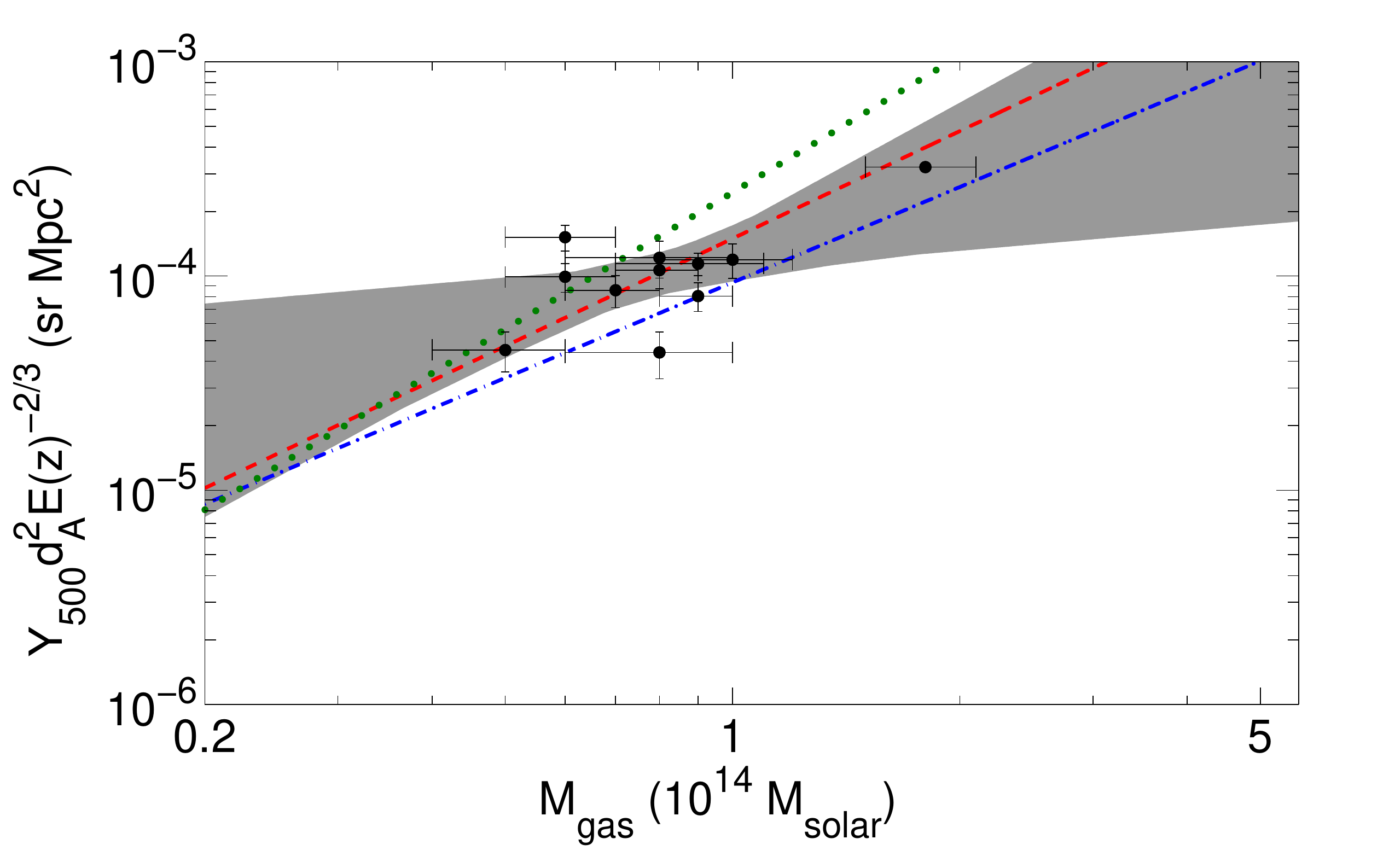}&
\includegraphics[width=0.32\textwidth]{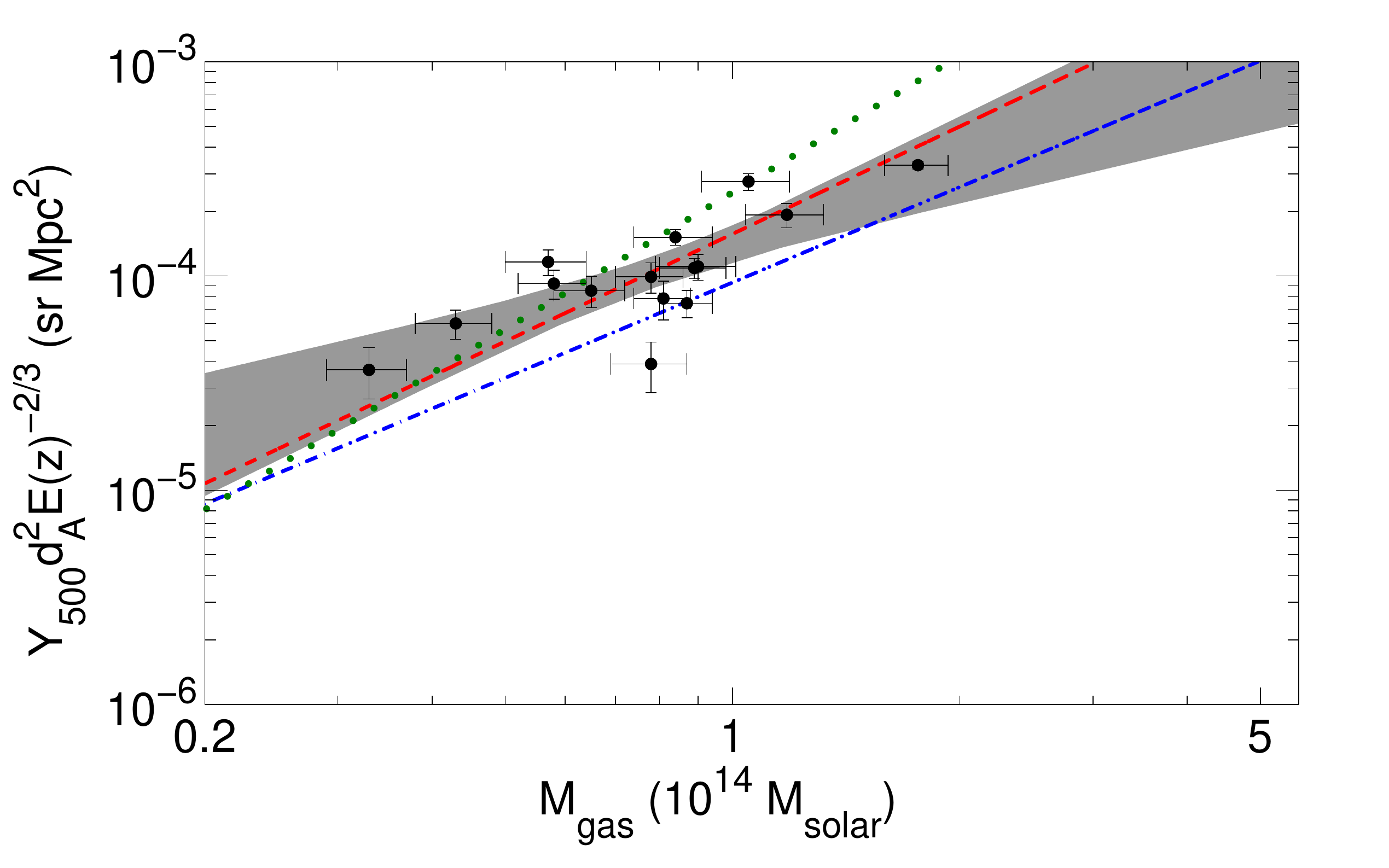}&
\includegraphics[width=0.32\textwidth]{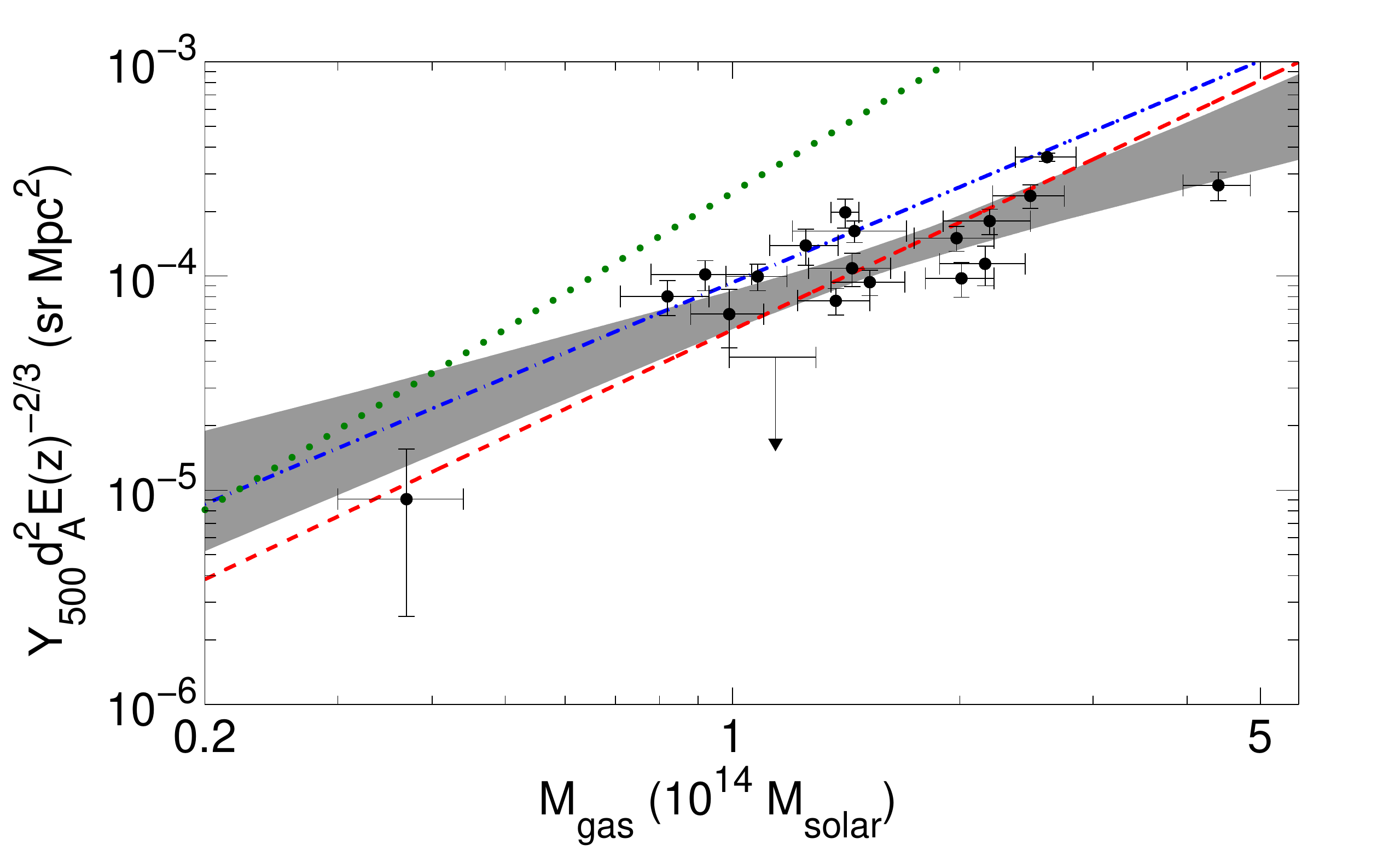}\\
\includegraphics[width=0.32\textwidth]{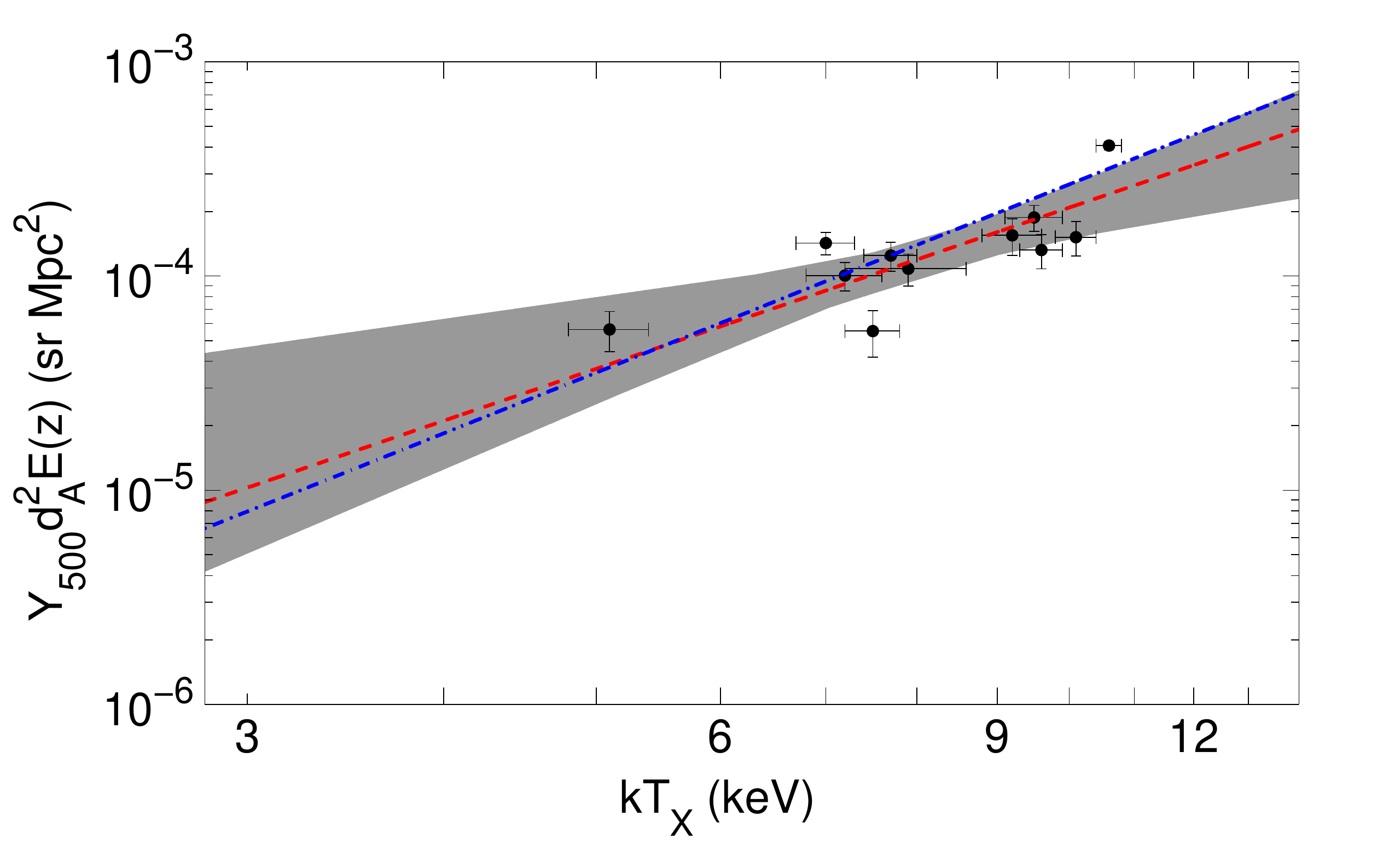} &
\includegraphics[width=0.32\textwidth]{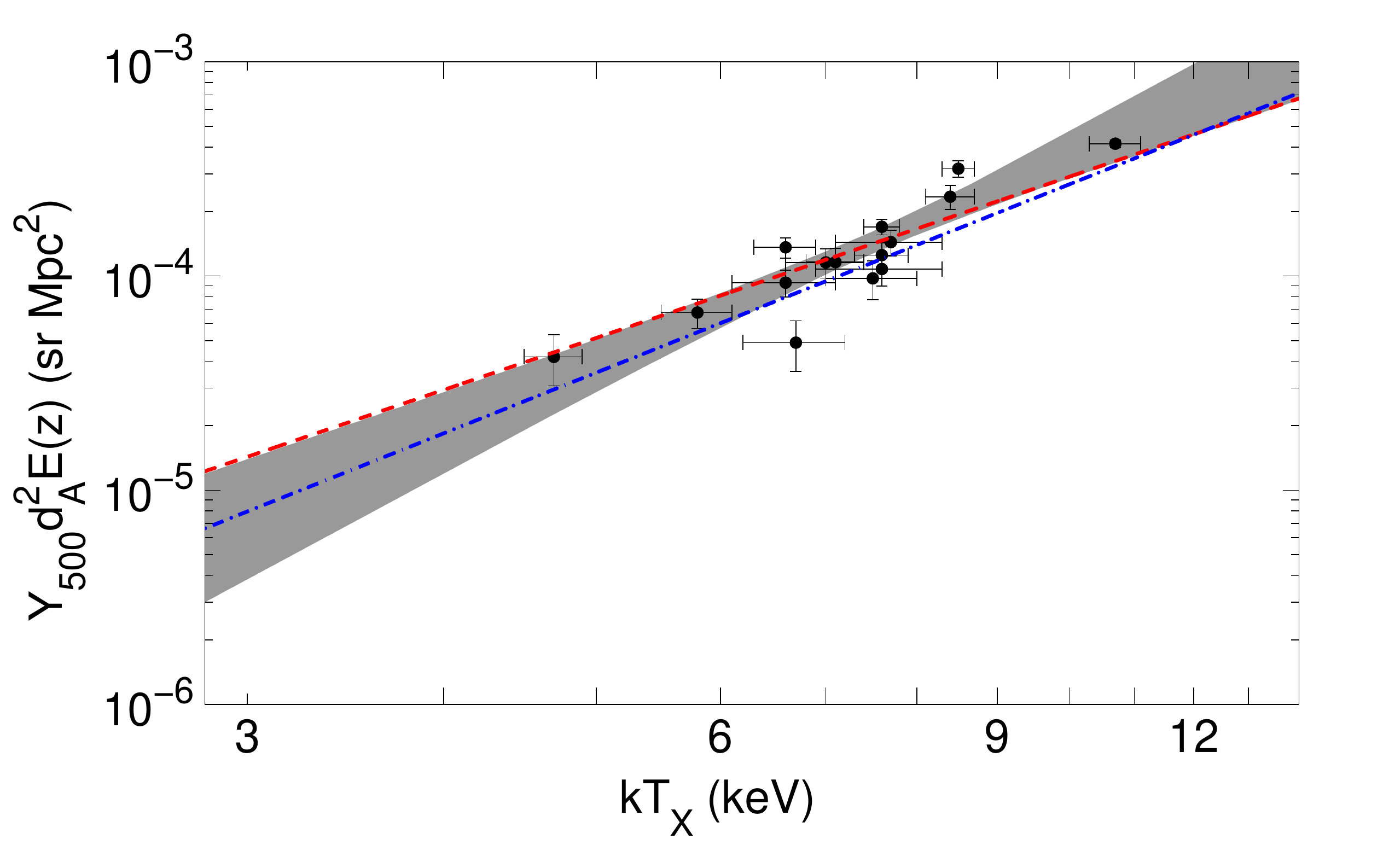} &
\includegraphics[width=0.32\textwidth]{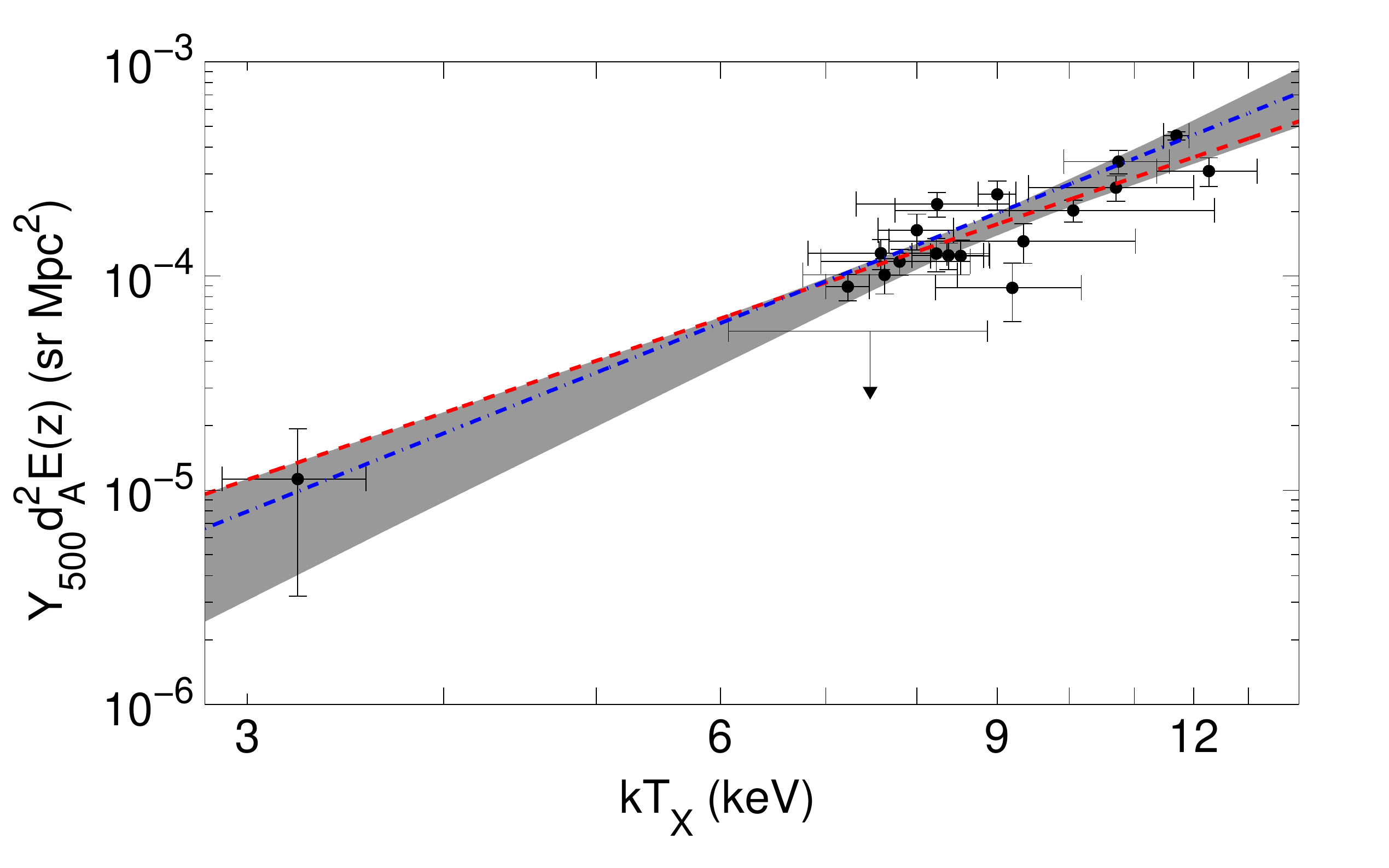} \\
\end{tabular}
\caption[]{$Y_{500}$ - $Y_X$, $Y_{500}$ - $M_{\rm{gas}}$, and $Y_{500}$ - $T_X$ scaling relations for the REFLEX-DXL (left column) and Z08 (middle column) and M10 (right column) cluster samples.  The black points show the APEX-SZ measured values given in Table \ref{TBL:mcmc}.  Non-detections are plotted as 95 per cent upper limits. The shaded region depicts the 68 per cent confidence interval of the power law for the three-parameter fit.  Also shown is the maximum likelihood result for the case where the exponent is fixed to the self-similar value (red dashed).  For comparison, we plot the best-fit results from \protect\citet{planck2011XI} (blue dot-dash), \protect\citet{plagge2010} (dark green dot), and \protect\citet{andersson2011} (light green dot).}
\label{fig:scalefullbeta}
\end{figure*}

To derive best-fit scaling relations parameters we perform a regression analysis of the data shown in Figure \ref{fig:scalefullbeta}.  We assume that the data follow the power law relationship, 
\begin{equation}
\label{EQN:powlaw}
Y_{500} d_A^2E(z)^{\delta} = A \cdot \left(\frac{ X}{X_0}\right)^B,
\end{equation}
where $A$ is the normalization, $B$ is the power law exponent, and $X$ denotes an X-ray observable.  A pivot point, $X_0$, is chosen for each observable ($T_0$ = 7.5 keV, $M_{\rm{gas},0} = 1.0 \times 10^{14} M_\odot$, $Y_{X,0} = 8.0 \times 10^{14} M_\odot \rm{keV}$) to reduce the correlations between $A$ and $B$.  We use the self-similar values of $\delta = (1, -2/3, 0)$, respectively (see Equation \ref{EQN:scale}).  The intrinsic astrophysical scatter in the scaling relation is represented through a log-normal distribution with mean $\mu = \rm{log}_{10}A +B\cdot \rm{log}_{10}(X/X_0)$ and variance $\sigma_{\rm{log_{10}(Y)}}^2$ in the base 10 logarithm of the distribution.

This regression is often performed in the log basis, where Equation \ref{EQN:powlaw} transforms to a simple linear relationship with Gaussian intrinsic scatter.  In agreement with \citet[][hereafter K07]{kelly2007}, we find from simulated data sets that common methods for linear regression with uncertainty in both the x- and y-variables, such as those in \citet{press2007} and \citet{weiner2006}, introduce systematic bias into the measured parameters.  The regression parameters are further biased by the symmetrization and transformation of the $Y_{500}$ likelihood distribution from the linear basis to the logarithmic basis.   Lastly, the measured $Y_{500}$ non-detections in the APEX-SZ data cannot be fully represented in the log basis.  For example, when we place a prior that $Y_{500} > 0$ and perform a  linear regression using the Bayesian method of sampling the posterior probability from  K07, we find that the resulting maximum likelihood parameters are biased at the $1-2\sigma$ level.

To address the difficulties introduced by the log basis, we instead perform the regression analysis in the linear basis.   We construct the likelihood of the scaling relations parameters from the probability densities of \y500 measured with APEX-SZ and the X-ray observable (e.g., $Y_X$).  
We approximate the measured probabilities in $Y_{500}$ as asymmetric Gaussian distributions. 
Following the methodology of K07, we model the distribution of $X$ as a weighted sum of Gaussian functions.\footnote{The MCMC algorithm is largely based on the publicly available IDL code from \citet{kelly2007} which can be found at: http://idlastro.gsfc.nasa.gov/ftp/pro/math/linmix\_err.pro.}  
 We sample the posterior likelihood distribution using an MCMC method with a Metropolis-Hastings algorithm.  Using simulated data sets with X-ray properties based on those of the APEX-SZ cluster sample, we find that we are able to recover unbiased estimates of the scaling relation parameters.  

\subsection{$ Y_{500}-Y_X, Y_{500}-M_{\rm{gas}}, Y_{500}-T_X$ Relations}
\label{SEC:sampleintro}

We fit the three scaling relations described in the previous section for three different subsamples within the APEX-SZ cluster sample.  These subsamples are created based on the different  X-ray analyses available: the REFLEX-DXL clusters, clusters from the analysis of Z08, and clusters from the analysis of M10.  By restricting our scaling relations to subsamples with uniform X-ray analysis, we avoid the systematic bias and inflated uncertainty that results from different analysis choices (see Section \ref{SEC:nonuniform}).  Several clusters are common between these three samples.  For these cases, the X-ray parameters from the desired sample are used when modelling each cluster and fitting the scaling relation.

APEX-SZ observed 11 of the 13 clusters from the X-ray selected REFLEX-DXL cluster sample\footnote{Technically, the REFLEX-DXL sample also includes RXCJ2011.3-5725, which was observed by APEX-SZ.  However, \citet{zhang2006} exclude this cluster from their detailed analysis due to flare contaminated observations.  Therefore, we also exclude it from REFLEX-DXL subsample to maintain a uniform X-ray analysis.} \citep{zhang2006}.  The remaining two clusters were not observed  due to restrictions in accessible sky area.  These clusters have an X-ray luminosity $L_X > 5.9 \times 10^{44}$ ergs\,$\rm{s}^{-1}$ in the ROSAT-ESO Flux Limited X-ray (REFLEX) galaxy cluster survey \citep{bohringer2004} and fall within a narrow redshift range, $0.27 < z < 0.31$. \citet{zhang2006} observed the REFLEX-DXL clusters with the \textit{XMM-Newton} satellite and measured the X-ray temperature and gas mass for each.  

The second sample for which we fit the scaling relations includes 15 of the 37 clusters from the Local Cluster Substructure Survey (LoCuSS\footnote{http://www.sr.bham.ac.uk/locuss}) measured by Z08.    These X-ray luminous clusters occupy a wider redshift range $0.14 \leq z \leq 0.3$ than the REFLEX-DXL sample.  Using observations from the \textit{XMM-Newton} satellite, these authors modelled the X-ray emission for each cluster to measure the X-ray temperature and gas mass.

Finally, the APEX-SZ sample includes 19 of the 94  clusters observed  by M10.   In contrast to the two other subsamples, the M10 X-ray properties were measured using both the \textit{Chandra} and \textit{ROSAT} telescopes, with a cross-calibration between the two data sets.   Additionally, these clusters occupy a slightly larger range in redshift, $0.20 \leq z \leq 0.45$. 

Beyond the follow-up of the REFLEX-DXL subsample, many of the APEX-SZ clusters were selected in an \textit{ad hoc} manner.  We are unable to quantify the influence of this selection on the Z08 and M10 samples, resulting in an unknown level of systematic bias and uncertainty.

The maximum likelihood regression parameters and 68 per cent confidence intervals for all three samples are given in Table \ref{TBL:linefit} and shown in Figure \ref{fig:scalefullbeta}.   We also give results for the cases where we fix the power law exponent to the self-similar values.

\subsubsection{The REFLEX-DXL, Z08, \& M10 Samples}
\label{SEC:reflexdxl}

First, we consider our constraints on the power law exponent $B$ for all three scaling relations, shown in Table \ref{TBL:linefit}.   We find for the REFLEX-DXL sample that $B$  is consistent with self-similar expectations for all three scaling relations.  In contrast, for the Z08 sample \ytx relation, we find a marginally steeper power law than expected in the self-similar case and measured for the REFLEX-DXL sample.  We measure the exponent for the \ymgas relation to be consistent between the three samples, however, these values are marginally lower than the self-similar expectation.

\begin{figure}\centering
\begin{tabular}{c}
\includegraphics[width=0.5\textwidth]{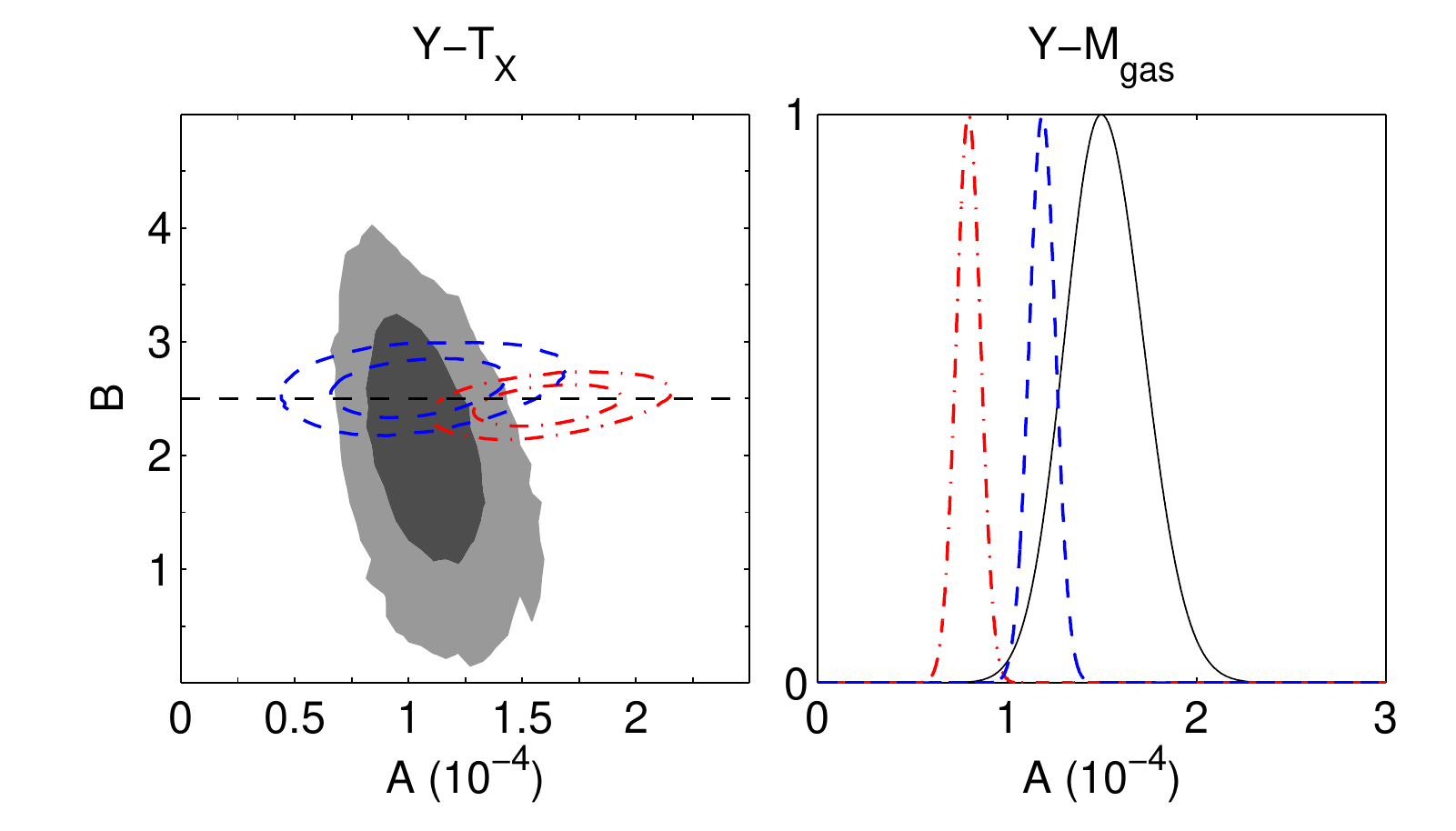}\\
\includegraphics[width=0.5\textwidth]{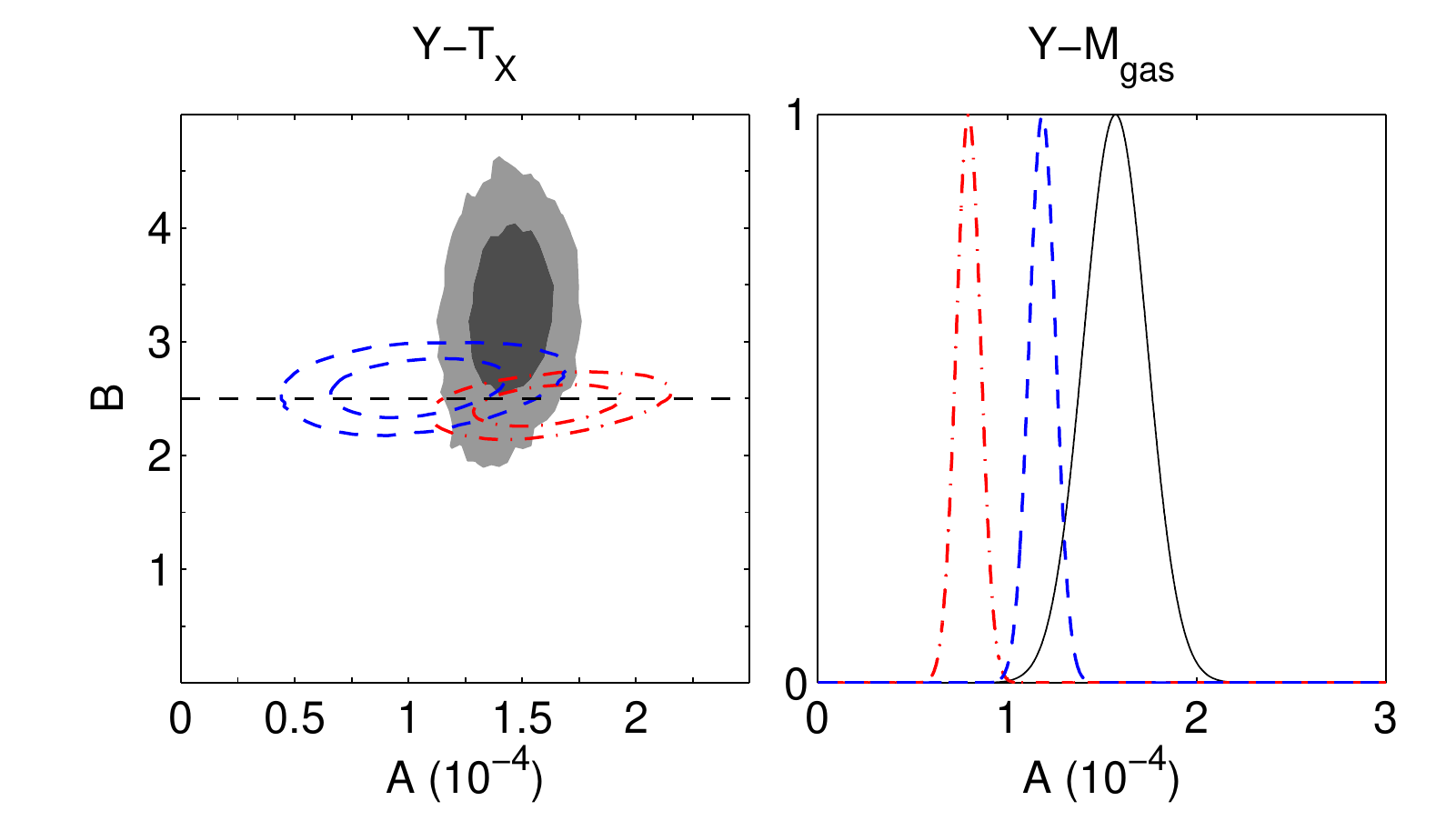}\\
\includegraphics[width=0.5\textwidth]{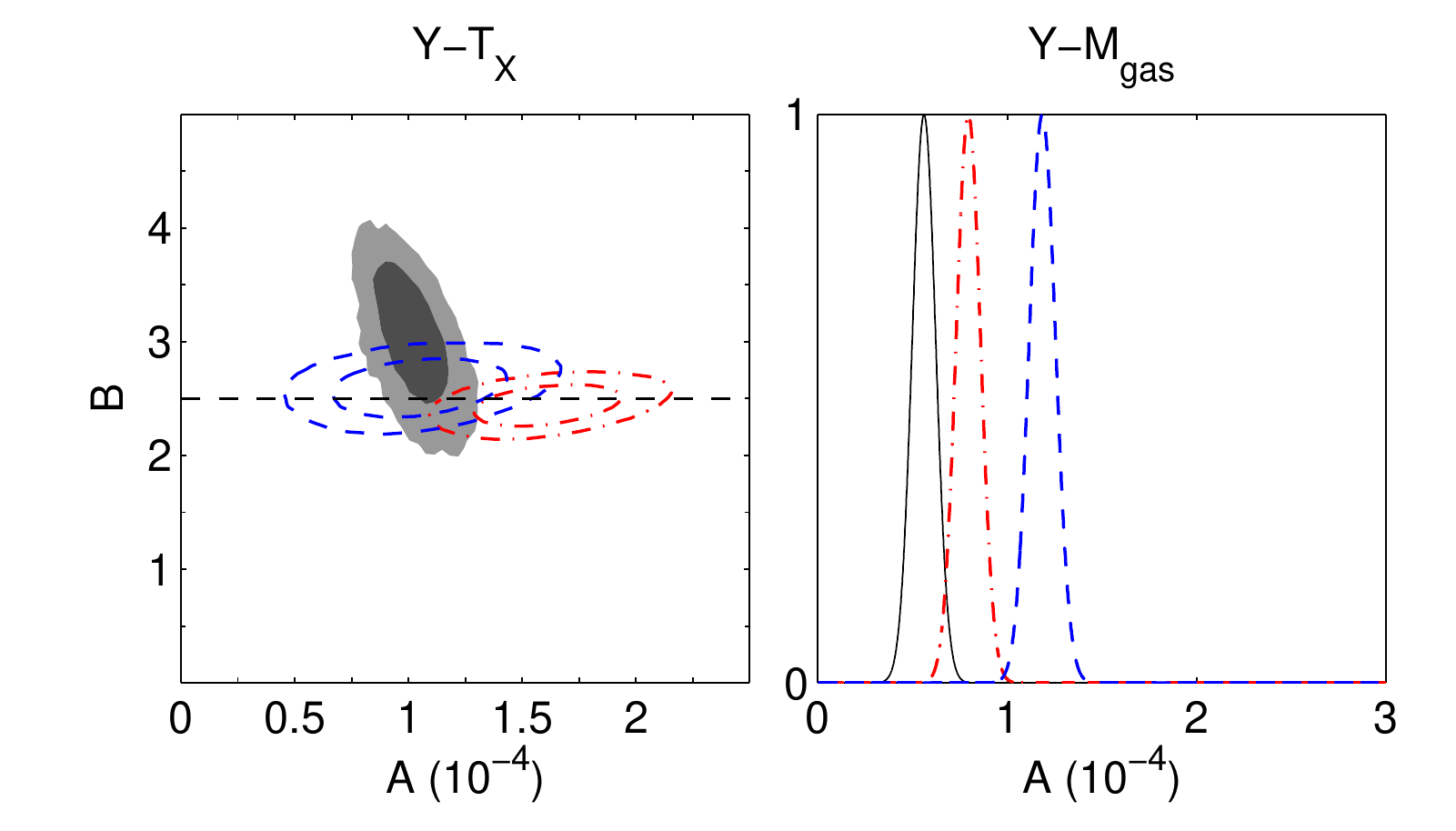}\\
\end{tabular}
\caption[]{Likelihood constraints for the REFLEX-DXL (top), Z08 (middle), and M10 (bottom) cluster samples.  The two-dimensional constraint for the normalization $A$ and power exponent $B$ is shown for the \ytx relation in the left panel.  The grey shaded areas represent the 68 per cent and 95 per cent  confidence regions and the dashed straight line shows the self-similar value of the exponent.  The \ymgas  constraint on $A$ for the case where the exponent is fixed to the self-similar value ($B = 5/3$) is shown in the right panel.  For comparison, the numerical results of \protect\citet{nagai2006} for standard gas dynamics only (red/dash-dot) and cooling and feedback (blue/dashed) are shown in both panels.}
\label{fig:ytxnagai}
\end{figure}

Next, we examine our measurements of the power law normalization $A$. For both the \yyx and \ymgas relations the M10 sample has a significantly lower $A$ (more than $3.3\sigma$ and $2.3\sigma$, respectively) than the other two samples.  The normalization of the \ytx relation is consistent for the REFLEX-DXL and M10 samples, while the Z08 sample suggests a higher value.  We compare our measurements for the $Y_{500}-M_{\rm{gas}}$ and $Y_{500}-T_X$ to the numerical simulations of \citet{nagai2006} (see Figure \ref{fig:ytxnagai}).   These simulations agree well with the self-similar model.  We therefore find similar tension  between the measured \ymgas exponent for all three samples and the simulations as for the comparison to self-similarity.  

Different models of cluster gas physics are expected to mostly influence the scaling relation normalization \citep{motl2005, nagai2006, lau2009}.  For this reason,  we consider the result where $B$ has been fixed to the self-similar value.  As shown in Figure \ref{fig:ytxnagai}, the \ytx results for all three samples and the \ymgas results for the REFLEX-DXL and Z08 samples favour simulations that include cooling and feedback over those with only standard gas dynamics.   In contrast, the lower normalization for the M10 sample \ymgas relation causes a preference for standard gas dynamics.   For all three samples, the \ymgas conclusions are weak because of the measured tension in the exponent.

The third free parameter in our regression model, $\sigma_{\rm{log_{10}(Y)}}$, quantifies the amount of scatter in the scaling relation not due to measurement uncertainties in $\rm{log_{10}}(Y_{500})$.   We transform the values of $\sigma_{\rm{log_{10}(Y)}}$ from Table \ref{TBL:linefit} into the fractional scatter, $\sigma_Y/Y=\sigma_{\rm{log_{10}(Y)}}/\rm{log}_{10}(e)$, commonly used in the literature.   We measure $\sim$28 per cent ($\sigma_{\rm{log_{10}(Y)}} \approx 0.12$) intrinsic scatter for all three samples.   \citet{kravtsov2006} demonstrated that $Y_X$ is a more robust mass proxy than $T_X$ or $M_{gas}$.  Because $Y$ and $Y_X$ measure similar quantities (Equation \ref{EQN:yyxscale}), we expect that \yyx will have the lowest intrinsic scatter of the three scaling relations.  As anticipated the \ymgas relation exhibits an increased level of intrinsic scatter (35 per cent).  The \ytx relation, however, has different values of intrinsic scatter  depnding on the sample (ranging from 14 per cent to 37 per cent).   We discuss these differences  further in Section \ref{SEC:systematics}.

\subsubsection{Comparison with Previous Experiments}
There are four main observational studies with which we directly compare our \y500 scaling relations.  \citet[][hereafter P11]{planck2011XI} and \citet[][hereafter P13]{planck2013xxix} present SZE scaling relations using the Planck satellite (the early and 2013 results, respectively) in combination with X-ray data using the \textit{XMM-Newton} satellite.   \citet[][S10]{plagge2010} and \citet[][S11]{andersson2011} constrain SZE scaling relations using the South Pole telescope.  Results from these studies are compared to the APEX-SZ results in Figure \ref{fig:scalefullbeta}.  

The APEX-SZ results for all three samples agree well with both the P11 and P13 constraints on the \ymgas and \yyx  exponents. S11 explore \yyx and likewise measure an exponent consistent with the APEX-SZ result. 
S10 evaluate both \ymgas and \yyx and measure exponents which are higher than the APEX-SZ values (approximately $1.6\sigma$ for the most consistent sample, REFLEX-DXL).  However, these authors note that their results vary significantly when using different values of gas mass from the literature.

To compare the measured APEX-SZ power law  normalization of the \yyx relation for the REFLEX-DXL sample (given in Table \ref{TBL:linefit}) to previous measurements we transform our normalization parameter to $A' = A \times 10^{-4}/8\times10^{14} / C_{SZ}$, where $C_{SZ} = \sigma_{th}/m_ec^2\mu_em_p$ \citep{arnaud2010}.   In the case where $B=1$, the ratio of $A'=Y_{500}d_A^2 / Y_x$ is a function of the differences between the mass-weighted  and spectroscopic cluster temperatures.  

We find for all three subsamples that the \yyx normalizations are in some tension with previous observations.  The REFLEX-DXL ($A'=1.18_{-0.11}^{+0.21}$) and Z08 ($A'=1.41\pm0.15$) samples suggest higher normalizations than observations that indicate $A'$ is less than unity, including those from S11 ($A'=0.82 \pm 0.07$) and P13 ($A'=0.973\pm0.01$).   The M10 sample has a lower normalization  ($A' = 0.74_{-0.09}^{+0.07}$) that is also in tension with both the S11 and P13 measurements.

Similarly, we compare the APEX-SZ \ymgas and \ytx best-fit normalizations to those in P11 and S10.   For the REFLEX-DXL sample we measure a value of $A$ consistent with S10 and P11.  The normalization for the Z08 sample is slightly more discrepant compared to P11 ($\sim2\sigma$).  In comparison, the measured value of $A$ for the M10 sample is significantly  lower than the other two samples, and is $2.2\sigma (2.7\sigma)$ different from P11 (S10).   For \ytx, the APEX-SZ results for all three samples agree with that of P11.

The level of intrinsic scatter in the \yyx relation for the three samples presented here is similar to that observed by both P11 ($\sigma_{\rm{log_{10}(Y)}} =0.10 \pm 0.01$) and  S11 ($\sigma_{\rm{log_{10}(Y)}} =0.09 \pm 0.04$).   The P11 scatter in \ymgas is comparable to that of \yyx, while the scatter in \ytx increases.   The APEX-SZ results show a slight elevation in the level of intrinsic scatter that is not statistically significant.  The \ytx intrinsic scatter depends strongly on the cluster subsample, with the REFLEX-DXL sample showing a level consistent with P11.  The Z08 and M10 samples, however, have a lower level of intrinsic scatter ($\sim1.6\sigma$) than P11.

In general, we find that our results agree with previous measurements of SZE - X-ray scaling relations.  The one notable exception is the normalization of the \yyx relation.  For all three cluster samples, our \yyx results show some mild tension with both theoretical expectations and previous measurements.  Additionally, the individual subsamples show significant ($4\sigma$) differences in comparison to each other.  
\begin{center}
\begin{table*}
\begin{minipage}{140mm}
\begin{center}
\caption{Scaling Relation Regression Parameters}
\label{TBL:linefit}
\begin{tabular}{lcccccccc}
\hline\vspace{-3.5ex}\\
 & \multicolumn{4}{c}{Full Fit} & \multicolumn{4}{c}{Fixed Exponent} \\ 
Subset & $N_{\rm{clusters}}$&A &B  & $\sigma_{\rm{Y,int}}$ & $N_{\rm{clusters}}$& A &B  & $\sigma_{\rm{Y,int}}$\\
\hline\vspace{-2.5ex}\\
 &\multicolumn{8}{c}{$Y-Y_X$}\\
\hline\vspace{-2.5ex}\\
REFLEX-DXL&11&$1.34_{-0.13}^{+0.24}$&$1.06_{-0.32}^{+0.24}$&$0.11_{-0.05}^{+0.07}$&11&$1.41_{-0.16}^{+0.16}$&$1.00$&$0.13_{-0.06}^{+0.03}$\\
Z08&15&$1.60_{-0.17}^{+0.17}$&$1.01_{-0.19}^{+0.17}$&$0.12_{-0.03}^{+0.05}$&15&$1.59_{-0.13}^{+0.16}$&$1.00$&$0.13_{-0.04}^{+0.03}$\\
M10&19&$0.84_{-0.10}^{+0.08}$&$0.88_{-0.11}^{+0.16}$&$0.12_{-0.03}^{+0.04}$&19&$0.80_{-0.07}^{+0.06}$&$1.00$&$0.11_{-0.03}^{+0.04}$\\
\hline\vspace{-2.5ex}\\
& \multicolumn{8}{c}{$Y-M_{\rm{gas}}$}\\
\hline\vspace{-2.5ex}\\
REFLEX-DXL&11&$1.35_{-0.27}^{+0.22}$&$1.07_{-0.43}^{+0.52}$&$0.15_{-0.05}^{+0.08}$&11&$1.50_{-0.19}^{+0.22}$&$1.67$&$0.13_{-0.06}^{+0.03}$\\
Z08&15&$1.40_{-0.15}^{+0.23}$&$1.32_{-0.34}^{+0.27}$&$0.15_{-0.04}^{+0.05}$&15&$1.57_{-0.17}^{+0.17}$&$1.67$&$0.13_{-0.03}^{+0.05}$\\
M10&19&$0.70_{-0.10}^{+0.10}$&$1.16_{-0.19}^{+0.25}$&$0.15_{-0.03}^{+0.05}$&19&$0.56_{-0.06}^{+0.06}$&$1.67$&$0.16_{-0.02}^{+0.04}$\\
\hline\vspace{-2.5ex}\\
& \multicolumn{8}{c}{$Y-T_X$}\\
\hline\vspace{-2.5ex}\\
REFLEX-DXL&11&$1.03_{-0.13}^{+0.19}$&$2.14_{-0.78}^{+0.60}$&$0.16_{-0.04}^{+0.06}$&11&$1.02_{-0.13}^{+0.14}$&$2.50$&$0.16_{-0.05}^{+0.05}$\\
Z08&15&$1.45_{-0.12}^{+0.12}$&$3.30_{-0.51}^{+0.46}$&$0.09_{-0.04}^{+0.04}$&15&$1.42_{-0.13}^{+0.12}$&$2.50$&$0.10_{-0.02}^{+0.05}$\\
M10&19&$1.01_{-0.11}^{+0.10}$&$3.04_{-0.37}^{+0.44}$&$0.06_{-0.03}^{+0.04}$&19&$1.11_{-0.10}^{+0.08}$&$2.50$&$0.07_{-0.01}^{+0.05}$\\
\hline\vspace{-2.5ex}\\
\end{tabular}
\end{center}
The values of $A$ presented in this table are in units of $10^{-4}$.
\end{minipage}
\end{table*}
\end{center}

\section{Systematic Uncertainties}
\label{SEC:systematics}
We divide our investigation of possible systematic contamination of these results into two categories: instrumental and analysis systematics and astrophysical effects.   In the following two sections we discuss these systematics including the role of sample selection,  correlations between the X-ray and SZE measurements, variations in X-ray measurements, known clusters that APEX-SZ did not detect, the assumptions inherent in the cluster profile modelling, contamination due to point sources, and a non-universal gas mass fraction.
\subsection{Instrumental and Analysis Uncertainties}

\subsubsection{Correlated Scatter}
A portion of the measured intrinsic scatter in $Y_{500}-T_X$ and $Y_{500} -Y_X$ can be attributed to correlation introduced by the fact that we use the X-ray temperature $T_X$ to estimate $r_{500}$ and therefore \y500. 
We measure this correlation in the scaling relations by simulating the APEX-SZ dataset.  For each cluster, a new temperature is drawn from its measured probability distribution and $r_{500}$ is determined from the $r_{500}-T$ relation.  The $Y_{500}$ distributions are recalculated and the regression analysis is performed on each realization of the scaling relation.   We find that this correlation accounts for an intrinsic scatter of  $\sim$6 per cent in the scaling relations,  which is less than half the total measured scatter for the subsample with the lowest value (M10).

\subsubsection{Non-Uniform X-ray Analysis}
\label{SEC:nonuniform}

The X-ray observables used in the APEX-SZ scaling relations are drawn from several different pre-existing studies in the literature.  We select the clusters in each of our scaling relations subsamples to ensure that a uniform X-ray analysis is used for each.  However, differences between X-ray analyses translate into a systematic bias in the $Y_{500}$ scaling relations for each sample.

\citet{plagge2010} used X-ray observables from several different studies and found that the measured \ymgas, and by extension, \yyx relations changed significantly depending on the choice of study.  \citet{rozo2012A} compared hydrostatic mass estimates of common clusters in three different X-ray samples.  These authors found total mass differences as large at 45 per cent at a redshift of 0.2, resulting in differences in $r_{500}$ and the aperture used to measure $M_{\rm{gas}}$ and $T_X$.   These aperture differences are in addition to those regarding the specific fraction of $r_{500}$  (e.g., $0.15 - 1r_{500}$) used by individual authors for $T_X$ measurements (see the note in Table \ref{TBL:physparam}).  For cool-core clusters in particular the core region is known to bias estimates of the cluster temperature and is often excluded, but the exact boundary chosen varies between authors.  Additionally, the outer radius is often determined by data quality limitations.    

\citet{rozo2012c} measured $Y_X$ from \textit{Chandra} observations for a subset of the clusters in \citet{planck2011XI} and compared these data with the \textit{Planck} SZE measurements to constrain the \yyx relation.  These authors found a significantly lower level of  intrinsic scatter  in comparison with the original \textit{Planck} + \textit{XMM-Newton} analysis of the full sample ($8.2\pm 3.5$ per cent and $22.8 \pm 2.3$ per cent, respectively).  When considering the same subset of clusters as the \textit{Chandra} analysis, \citet{rozo2012c} measure a $16.7 \pm3.9$ per cent intrinsic scatter.  \citet{rozo2014b} extends this analysis by comparing \textit{Planck} data \citep{planck2011XI} with $Y_X$  from \citet{mantz2010b} and suggest an upper limit on the intrinsic scatter of 15 per cent.  These authors conclude that the observed variations are the result of systematic differences in the X-ray analysis.

We investigate the importance of uniform X-ray analysis by fitting scaling relations to the full APEX-SZ sample of 42 clusters and comparing the results to the subsets in Table \ref{TBL:linefit}.   For all three scaling relations, we find the measured intrinsic scatter for a uniformly analyzed subset is less than when considering the full sample.  The most dramatic improvement in scatter is seen for the \citet{mantz2010b} sample,  decreasing for the \ytx relation from 41 per cent to 14 per cent.  By drawing random sets of clusters from the full sample and performing the regression analysis, we conclude that this level of scatter has a 1 per cent chance of occurring based on the parent distribution and that the observed decrease in intrinsic scatter is significant.  

As discussed in Section \ref{SEC:reflexdxl}, we find that there are significant variations in both the normalization and exponent for the \ytx relation between the three independent subsamples.  The \ymgas relation has less variation in the exponent, while the normalization changes by a factor of two between the \citet{zhang2008} and \citet{mantz2010b} subsamples.  The \yyx relation shows a similar factor of two variation in the normalization.   Overall, we find that using X-ray parameters from multiple studies introduces significant variability and uncertainty in the power law parameters due to systematic analysis differences.  Additionally, the level of intrinsic scatter is dominated by systematics when using multiple studies. This increase is expected as the intrinsic scatter attempts to account for the difference in normalizations.


\subsubsection{Cluster Non-Detections}

There are seven clusters in the full sample that APEX-SZ did not detect (defined by $Y_{500} < \sigma_{Y_{500}}$).  These clusters are specified in Table \ref{TBL:mcmc}. Included in the non-detections are the three clusters with the highest redshifts: RDCSJ1252-2927 ($z = 1.240$), XMMXCSJ2215.9-1738 ($z = 1.45$), and XMMUJ2235.3-2557 ($z = 1.393$).   These clusters were also observed at 31 GHz by \citet{culverhouse2010} using the Sunyaev-Zel'dovich Array (SZA).   Of the three, only XMMUJ2235.3-2557 was detected by SZA.  The APEX-SZ $Y_{500}$ likelihood distributions for RDCSJ1252-2927 and XMMXCSJ2215.9-1738 are consistent with the SZA upper limits.  \citet{culverhouse2010}  conclude that RDCSJ1252-2927 and XMMXCSJ2215.9-1738 are lower mass systems than expected from X-ray measurements. The seven non-detections are not part of the REFLEX-DXL or Z08 samples, and therefore will not impact the results presented in Table \ref{TBL:linefit}.  The M10 sample contains a single non-detection.  We investigate the influence of this cluster by comparing the measured scaling relation regression parameters with and without it included and find no signficant difference.

\subsubsection{Pressure Profile Model}
A final analysis systematic to consider is the assumption that the \citet{arnaud2010} model with our inferred values of $r_{500}$ accurately represents the true cluster pressure profile.  \citet{andersson2011} find that \y500 varies by $\sim10$ per cent under a variety of model assumptions, including the \citet{arnaud2010} profile.   The limited angular extent and strong timestream filtering of the APEX-SZ data prevent us from exploring different model parametrizations.  However, we list the PTE for the \citet{arnaud2010} model fit to each cluster in Table \ref{TBL:mcmc}.   If we exclude clusters which have a poor fit to the chosen \citet{arnaud2010} model ( $\sim25$ per cent of the sample with PTE $< 1$ per cent or PTE $>99$ per cent), the power law regression parameters do not change significantly.

\subsection{Astrophysical Uncertainties}
\subsubsection{Point Sources}
At the APEX-SZ observing frequency, emission from point sources can mask the SZE signal from a cluster. 
We estimate the effect of radio sources by extrapolating the flux densities of sources in the NRAO VLA Sky Survey  \citep[1.4 GHz, ][]{condon1998}, VLA FIRST \citep[1.4 GHz, ][]{becker1995}, and Parkes-MIT-NRAO \citep[4.85 GHz, ][]{griffith1993} surveys to 150 GHz.   Assuming a power law spectral energy distribution,  $S \sim \nu^\alpha$, with $\alpha =- 0.7$, there are six clusters with known bright sources with $\Delta T_{\rm{CMB}} > 100\,\mu$K within two arcmin of the X-ray centroid.   Of these six, two (RXCJ2014.8-2430 and MACSJ1931.8-2635) are not detected by APEX-SZ.   

Simulations from \citet{sehgal2010} suggest that only 1 per cent of clusters with masses similar to those observed in this paper will have radio sources at 150 GHz that contaminate the integrated cluster signal by 20 per cent or more.   For the APEX-SZ clusters, this implies that a single cluster would suffer from radio contamination and that our estimate of six contaminated clusters is extremely conservative.  

We have also observed a subset of the APEX-SZ clusters with the CARMA array, searching for point source contamination.  Nine clusters were observed at 100 GHz and three at 230 GHz with an average map rms of approximately 0.9 and  4.4 mJy\,$\rm{beam}^{-1}$, respectively.  We detect a 7.4 mJy  source at 100 GHz in the map of RXCJ1504.1-0248. 
Of the three non-detection clusters observed with CARMA (XMMUJ2235.3-2557, XMMXCSJ2215.9-1738, XMMUJ1230.3+1339) no point sources are detected. 

We test for the effects of these bright point sources by excluding the associated clusters from the appropriate subsamples (five clusters for the M10 sample, and one each for the REFLEX-DXL and Z08 samples).  There is no significant difference in the measured scaling relations parameters when removing these potentially contaminated clusters.

\subsubsection{Non-Universal Gas Mass Fraction}
The final systematic we investigate is a gas mass fraction that changes as a function of total mass.  There is evidence in both observations  \citep{zhang2006,vikhlinin2009b} and simulations \citep{battaglia2013} that the gas mass fraction changes due to the thermodynamics and feedback processes in the intracluster medium.  As seen in Equation \ref{EQN:scale}, both the \ytx and \ymgas relation depend on the gas mass fraction.  All of the results presented in Table \ref{TBL:linefit} assume a constant gas mass fraction embedded in the normalization term.  We test the effect of a non-universal gas mass fraction by first estimating cluster mass using a scaling relation with  $T_X$ from \citet{vikhlinin2009b}.  We adopt the functional form for gas mass fraction given  by the same authors, $f_{\rm{gas}}(h/0.72)^{1.5} = 0.125 +0.037\cdot\rm{log}(M_{15})$, where $M_{15}$ is the total cluster mass $M_{500}$ in units of $10^{15} h^{-1}M_\odot$.  We then look for the effect of changing $f_{\rm{gas}}$ on the power law exponent parameter in the measured scaling relations. 

When accounting for the non-universal $f_{\rm{gas}}$ we find a \ymgas power law exponent of $B = 1.34_{-0.55}^{+0.45}$, $B = 1.43_{-0.28}^{+0.31}$, and $B = 1.31_{-0.24}^{+0.24}$  for the REFLEX-DXL, Z08, and M10 samples, respectively.  These values are consistent with the results where we assume a constant $f_{\rm{gas}}$, and are consistent with self-similarity.  
In addition, the intrinsic scatter increases from $\sigma_{\rm{log_{10}(Y)}} = 0.15$ to $\sigma_{\rm{log_{10}(Y)}} = 0.19$ for the REFLEX-DXL and M10 samples. However, the stated uncertainty in the $f_{\rm{gas}}$ relation is $\sim4-5$ per cent \citep{vikhlinin2009} and our use of $T_X$ combined with the scatter already present in the \ymgas relation most likely explains the increase.

In summary, we examine the role of millimeter-wavelength point sources and a non-universal gas mass fraction as astrophysical systematics in the APEX-SZ scaling relations.  We find no evidence for bias or additional intrinsic scatter due to point sources.  Accounting for a non-universal gas mass fraction results in a slight change to the measured power law and increases the intrinsic scatter.

\section{Summary}
\label{SEC:conclusions} 
We present Sunyaev-Zel'dovich effect (SZE) observations of galaxy clusters measured with the APEX-SZ experiment and use them to investigate the scaling of the SZE with cluster mass.   We model the thermal pressure of the intracluster medium using the \citet{arnaud2010} universal pressure profile and use the results to calculate the spherical integrated Comptonization \y500 for each cluster.    We compare these measurements of \y500 to X-ray estimates of cluster mass taken from the literature.  Using these two sets of observables, we measure the \yyx, \ymgas, and \ytx scaling relations, finding the best-fit power law and intrinsic scatter for each.    These scaling relations are measured for three subsamples of the 42 APEX-SZ clusters that have uniform X-ray analysis: the REFLEX-DXL sample, 15 clusters from Z08, and 19 clusters from M10.   We find when all 42 clusters with varying X-ray analyses are included, significant systematics are introduced into the power law regression parameters and the intrinsic scatter increases.

For all three subsamples, we find that the best-fit power laws for the \yyx, \ymgas, and \ytx relations have exponents consistent with those predicted by the self-similar model.  We compare the measured normalizations for each scaling relation to numerical simulations to probe the underlying astrophysics of the intracluster medium (ICM).  The measured normalization of the \yyx relation for two of the subsamples implies a higher $Y_{500}d_A^2/Y_X$ than seen in previous studies.  However, there is a large variation in the normalization between the three samples, despite the uniformly analyzed X-ray data.  Therefore, we associate a large systematic uncertainty with the high $Y_{500}d_A^2/Y_X$ and do not draw further conclusions from it.   We compare the normalization for the \ymgas and \ytx relations to the numerical simulations of \citet{nagai2006} and find a weak preference for models that included radiative cooling and feedback in the ICM as well as standard gas dynamics.

Finally, we find that the levels of intrinsic scatter for the \yyx and \ymgas relations are consistent with previous measurements. 
The uncertainties in the intrinsic scatter are large due to the small number of clusters in the three subsamples.  A larger sample is key to improving this measurement of intrinsic scatter in \y500 and its application as an estimator for total cluster mass in cluster-based cosmological constraints. 

A uniform X-ray analysis for the full sample of APEX-SZ  clusters is ongoing.  This analysis will resolve the main systematic limitation of the full cluster sample and will allow us to use its large number of clusters to improve on the constraints presented here for the three subsamples.  Additionally, a follow-up program of optical observations has been concluded to estimate the total cluster mass independently using  weak-lensing measurements.  Future papers will use the additional X-ray and weak-lensing information to improve further our understanding of how \y500 scales with total cluster mass and facilitate its use in exploring the physics of the ICM as well as in constraining cosmological parameters using galaxy clusters. 

\section*{Acknowledgements}
We thank the staff at the APEX telescope site, at the time of observations led by David Rabanus and previously by Lars-\AA ke Nyman, for their dedicated and exceptional support. We also thank Bradford Benson for useful comments on a draft version of the manuscript as well as LBNL engineers John Joseph and Chinh Vu for their work on the readout electronics.  APEX-SZ is funded by the National Science Foundation under Grant Nos.\ AST-0138348 \& AST-0709497.
Work at LBNL is supported by the Director, Office of Science, Office of High Energy and
Nuclear Physics, of the U.S. Department of Energy under Contract No. DE-AC02-05CH11231. Work at McGill is supported by the Natural Sciences and Engineering Research Council of Canada, the Canada Research Chairs program, and the Canadian Institute for Advanced Research.  This work has been partially supported by the DFG Transregio program TR33 "The Dark Universe".
FP acknowledges support from the BMBF/DLR grant Nr. 50 OR 1117.  CH acknowledges support from the Barbro Osher pro Suecia foundation and from the Swedish Research Council under grants 2006-3356 and 2009-4027. M.K. acknowledges support by the German BMWi through the Verbundforschung under grant 50\,OR\,1107.  NWH acknowledges support from the Alfred P. Sloan Foundation.


\end{document}